\documentclass[preprint2]{aastex6}

\usepackage{color}

\newcommand{\msun}{\ensuremath{\, {\rm M}_\odot}}

\newcommand{\rsun}{\ensuremath{\, {\rm R}_\odot}}
\newcommand{\lsim}{\mathrel{\hbox{\rlap{\lower.55ex \hbox {$\sim$}}
 \kern-.3em \raise.4ex \hbox{$<$}}}}
\newcommand{\gsim}{\mathrel{\hbox{\rlap{\lower.55ex \hbox {$\sim$}}
 \kern-.3em \raise.4ex \hbox{$>$}}}}

\begin{document}

\title{Electron Capture Supernovae from Close Binary Systems}
\author{Arend J.T. Poelarends, Scott Wurtz, James Tarka, Cole Adams and Spencer T. Hills}
\affil{Wheaton College\\
Department of Physics and Astronomy \\
501 College Ave \\
Wheaton, IL 60187 USA\\
E-mail: \texttt{aj.poelarends@wheaton.edu}}
%

\begin{abstract}
We present the first detailed study of the Electron Capture Supernova Channel (ECSN Channel) \added{for a primary star} in a close binary star system. Progenitors of ECSN occupy the lower end of the mass spectrum of supernovae progenitors and are thought to form the transition between white dwarfs progenitors\deleted{(lower initial masses)} and core collapse progenitors\deleted{(higher initial masses)}.\deleted{For single stars the mass range for ECSN is thought to be only $0.25 \msun$ wide and located at an initial stellar mass of $8-10\msun$.}The mass range for ECSN from close binary systems is thought to be wider \added{than the range for single stars}, because of the effects of mass \replaced{loss}{transfer} on the helium core. Using the MESA stellar evolution code \added{we explored the parameter space of initial primary masses between $8 \msun$ and $17 \msun$, using a large grid of models}\deleted{calculated a large grid of stellar models in the relevant parameter space various initial primary masses, secondary masses, periods and mass transfer efficiencies}. \replaced{We find that, in addition to the initial primary mass, the mass loss history is the most important factor in the final fate of stars in this mass range.}{We find that the initial primary mass and the mass transfer evolution are important factors in the final fate of stars in this mass range.} Mass \replaced{loss}{transfer due to Roche Lobe overflow} during and after carbon burning causes the core to cool down so that it avoids neon ignition, even in helium-free cores with masses up to $1.52\msun$, which in single stars would\deleted{certainly} ignite neon. \replaced{If the core is not able to recover from the effects of late mass \replaced{loss}{transfer}, it will continue to cool down and form a super-Chandrasekhar mass oxygen-neon white dwarf. However, if the core is able to reverse the downward temperature trend, and to contract to high enough densities for electron captures to commence,}{If the core is able to contract to high enough densities for electron captures to commence,} we find that, for the adopted Ledoux convection criterion, the initial mass range for the primary to evolve into an ECSN is between $13.5\msun$ and $17.6\msun$\deleted{with the range between $13.5\msun$ and $15.4\msun$ ($1.9\msun$ wide) consisting primarily of Case B systems and the range between $15.4\msun$ and $17.6\msun$ ($2.2\msun$ wide) consisting primarily of much less abundant Case A systems.}. The mass ratio, initial period, and mass loss efficiency only marginally affect the predicted ranges.

\end{abstract}
\keywords{Binaries: close  --- Stars: evolution --- Supernovae: general --- Methods: numerical}



\section{Introduction} \label{sec:intro}
Two types of supernova \added{explosions} are thought to be responsible for the creation of neutron stars \added{(NS)} in the universe. While the majority of these explosions are the result of a collapsing iron core after fuel exhaustion \citep[a so-called a core collapse supernova, CCSN, cf.][]{whw02,hfw+03,lan12}, a fraction of supernova progenitors most likely collapses as a result of the loss of pressure support due to electron captures on $^{24}$Mg and $^{20}$Ne \citep{nom84e}. These so-called electron capture supernovae (ECSN) are thought to occupy the lower end of the mass spectrum of supernova progenitors. ECSN thus form the transition between massive oxygen-neon (ONe) white dwarfs (WD) and supernovae. \replaced{As about $50\%$ of stars in this mass range}{As the majority of massive stars} are \replaced{thought}{observed} to be part of a binary system which could impact its evolution \citep{kf07,smk+12,skm+13,dk13,kkl+14, ast+17}, not only the observational properties of these stars will be very different from single stars \deleted{in this mass range}, \replaced{they}{their final properties} could also affect our understanding of the formation and evolution of NS+NS mergers which could be observed by the aLIGO/VIRGO network \citep{aaa+16, cbf+17}.

For single stars, \replaced{this}{the} transition region between WD and CCSN, has been explored quite thoroughly. Pioneering work was done in the eighties by \citet{mny+80, nom84e, nom87c, mn87}, followed by further work in the late nineties \citep{gi94, rgi96b, gri97b, irg97, rgi99}, and a final wave during the last ten years, some extending their models to lower metallicities  \citep{sie06, sie07, sie10, phl+07,tyu13, jhn+13, jhn14, dsl+10, dgl+14, dgl+14b, dgs+15}. The initial mass range, for which ECSN could occur in single stars, was initially predicted between $8 \msun$ and $10 \msun$, based on the mass of the helium core \citep{nom84e, nom87c}. This was later refined, for solar metallicity stars, to a much narrower range, especially due to a better understanding of the effect of the second dredge-up which reduces the helium cores of stars in the relevant mass range down to below the Chandrasekhar mass \citep{phl+07}. Uncertainties \replaced{regarding}{in} mass loss rates during the final phase of the evolution of stars in this mass range, combined with a lack of general consensus regarding the treatment of chemical mixing and convection in these stars, \added{however, }make that several estimates of the \added{initial} ECSN mass range now exist, ranging from $7-9 \msun$ \citep{wh15}, $9-9.25\msun$ \citep{phl+07} to $9.5-11\msun$ \citep{sie06, dsl+10, tyu13}. Fundamentally, however, the final fate of a star in this mass range is determined by a race between core growth and mass loss. If heating due to core growth is able to offset cooling due to neutrino losses the core will contract at a roughly constant temperature until a critical density is reached where electron captures can provide additional heating which will start O+Ne deflagration in the very center, leading to an explosion \citep{tyu13}. However, if the star experiences a strong stellar wind at the end of its life (not unlikely for stars on the Asymptotic Giant Branch), \replaced{it could be possible that it will lose its envelope before conditions are able to develop that are conducive for electron captures to start, and its final fate will be a massive ONe white dwarf}{it is possible that the entire envelop might be removed before the conditions necessary for electron captures are reached. In this scenario, the final fate would be a massive ONe WD} \citep{phl+07}.

\replaced{This}{The} ECSN mass range, however, is expected to be different in close binary systems as the primary \replaced{is expected to lose much}{could potentially lose a significant fraction} of its mass as a result of Roche lobe overflow \citep[hereafter RLOF, cf.][]{wlb01, lan12, mli+13}. \citet{plp+04} speculated that \deleted{the mass range for}ECSN could \replaced{be expanded possibly to $8-17 \msun$}{occur for primary masses between 8\msun\ and 17\msun}. They argued that \deleted{for Case A systems (} systems that start mass transfer during core hydrogen burning\deleted{)}\added{(Case A)} \deleted{this} would \replaced{result in a}{give rise to} much smaller \added{helium} cores, pushing the limit for ECSN toward higher initial masses. \deleted{For Case B systems (}Systems that start mass transfer during or after hydrogen shell burning\deleted{)}\added{(Case B)} \added{would avoid the reduction of the helium core by the second dredge-up}\deleted{the helium core mass would not be reduced by the second dredge-up (because of the effects of mass \replaced{loss}{transfer})} and\deleted{this} could potentially \replaced{result in}{form} a bigger \added{helium} core, thus \replaced{extending}{expanding} the range also to lower initial masses. 

Attention has recently turned to stripped-envelope stars in close binaries as possible progenitors of ECSN \citep{tlm+13, tlp15,me16}. Motivated by recent discoveries of weak and fast optical transients, ECSN from close binaries have been suggested as a possible origin, due to the fact that through binary interactions, ECSN progenitors can lose most or all of their hydrogen envelope. \citet{tlm+13, tlp15} showed that a helium star companion to a neutron star may experience mass transfer and evolve into a ONe core with a mass of $\sim 1.5 \msun$ which in certain binary configurations may lead to an ECSN. \citet{me16} explore the possibility of ECSN from mergers and the effects of common envelope evolution. They show that binaries with short orbital periods and fairly high mass ratios are able to experience a common envelope (CE) phase and a subsequent merger, after which the product of the merger is able evolve into an ECSN with a small amount of ejecta due to the significant mass loss during the CE phase and merger.

In this paper we explore the parameter space established by \citet{plp+04}\added{, primarily defined by primary masses between 8\msun\ and 17\msun,} to more accurately determine whether ECSN from close binary systems are indeed a possibility and which mass range they would occupy. The predictions of \citet{plp+04}\added{, however,} \replaced{are}{were} based on the helium core criterion as defined by \citet{nom84e,nom87c}\deleted{ which suggested an ECSN mass range in single stars of $8 - 10 \msun$}. \added{The validity of this criterion for binary systems, however, can be questioned as it is based upon the mass of the undisturbed helium core, and does not account for the effects of mass transfer.} \replaced{This}{In addition, the} mass range for single stars\deleted{, however,} has been narrowed down a bit in subsequent studies, leading to the question whether the $8 - 17 \msun$ mass range for ECSN in close binaries, as estimated by \citet{plp+04}, is still accurate. To investigate this, we created a large grid of models, covering most of the mass range suggested by \citet{plp+04}. \deleted{, four possible values of the mass ratio $q$ (0.6, 0.7, 0.8, 0.9), initial periods between $3$ and $65$ days, and four possible values of the mass transfer efficiency parameter $\beta$ (0.0 -- conservative, 0.25, 0.50, 0.75 -- mostly ineffective).}

In Section~\ref{sec:methods} we discuss our stellar evolution code, the input physics we employ, and the details of the grid. In Section~\ref{sec:rep_systems} we present several representative \deleted{Case A and Case B} systems and discuss the main impacts on the evolution of such systems. Section~\ref{sec:new_ECSN_mass_range} investigates the role of mass loss \added{due to Roche Lobe overflow}, \replaced{focusing on major mass loss episodes early in the evolution of the systems, during and after carbon burning, and in the final stages of the life of the primary.}{and the effects it has on the evolution of the core.} In Section~\ref{sect:new_ECSN_pathway} we explore various pathways for ECSN in binaries, based on detailed models of the final evolution of the carbon core. In Section~\ref{sec:ECSN_mass_range} we present updated values for the ECSN mass range in close binary systems. In Section~\ref{sec:discussion} we consider the question how probable ECSN from close binaries are, and what the expected mass range for ECSN would be, and we compare our results with previous work on this topic.

Appendix~\ref{sec:other_mixing} discusses the effects on our main results of different treatments of convective boundaries, and App.~\ref{sec:resolution} discusses the robustness of our results through a resolution study.

\section{Methods and Grids}\label{sec:methods}
\subsection{Stellar Evolution Code}
We used the MESA stellar evolution code \citep{mesa1, mesa2, mesa3} (revision 8118) to model the evolution of a dense grid of binary stellar evolution models.\deleted{With the enhancements described in \citet{mesa3} this code has now the capability to accurately model the evolution of stars in a binary system. Each star is evolved individually, while the binary interaction is provided by the \texttt{binary} module.} \added{Nuclear reactions were followed using the sufficiently detailed networks provided with MESA, i.e. \texttt{basic.net} for hydrogen and helium fusion, \texttt{co\_burn.net} for carbon and oxygen fusion, and \texttt{approx21.net} for later phases.} Opacities were calculated using tables from the OPAL project \citet{gn93}, with the initial metallicity set to $z = 0.02$ and metal fractions set according to \citet{gs98}.  Convection was treated according to standard mixing-length theory \citep{boh58} with a mixing-length parameter $\alpha = 1.5$ using the Ledoux criterion to determine the location of convective boundaries. \added{To maintain some degree of consistency with \citealt{wlb01}, no additional mixing due to overshooting was incorporated in the models,} and semi-convective mixing was modeled according to \citet{lef85} with $\alpha_\mathrm{sm} = 0.01$. \deleted{(similar to citealt{wlb01})} \added{This choice, particularly to leave out any effects of overshooting, does have implications for the growth of the core and the removal of the envelope, and will potentially affect our final ECSN mass ranges. It has been well established \citep{mae76b, sie07, mesa2} that using the Schwarzschild criterion for determining convective boundaries produces larger helium cores, especially when one includes a certain amount of overshooting. This will shift the ECSN mass range to lower initial primary masses. Overshooting also changes the response of the star to mass accretion and its ability to adjust its thermal structure, possibly leading to the secondary filling its Roche lobe at a different time, affecting the formation of contact in overshooting grids. We plan to perform a more comprehensive parameter study on the effects of convection criteria and overshooting in the future, but we have included a brief investigation into the effects of more effecient semi-convection, and the effects of overshooting in the context of the Schwarzschild criterion in Appendix~\ref{sec:other_mixing}.}

We modeled stellar winds according to the standard implementation in MESA (cool winds \& hot winds), following \citet{rei75} for stars with surface temperatures below $10,000$ K, and \citet{kpp87} for stars with surface temperatures above that. Mass transfer through Roche lobe overflow is calculated according to \citet{rit88} through \replaced{an}{the} implicit scheme described in \citet{mesa3}. MESA has the capability to handle accretion onto a critically rotating star, by either keeping the star at a set rotation rate (e.g. 98\% of critical rotation) and rejecting additional accreted matter \citep[cf.][]{mesa3,mlp+16} or by employing a scheme to enhance the mass loss at critical rotation \citep[cf.][]{hlw00}. However, as we \replaced{want}{aimed} to test the sensitivity of \replaced{these}{our} binary models to the mass transfer efficiency, we chose to set the mass transfer efficiency through Roche lobe overflow to fixed values. \deleted{As the main objective for this research was to establish whether the primary in a close binary system is able to evolve in to an ECSN (including outcomes across various mass ratios) we opted to}To avoid super-critical rotation of the secondary, we only follow the spin angular momentum of the primary; hence, we put the initial surface velocities at $100$ km\,s$^{-1}$ and $0$ km\,s$^{-1}$ for the donor and the accretor, respectively. \added{We model the tidal interactions on the primary as described in \citet{mesa3} with the synchronization timescale for convective envelopes calculated according to \citet{htp02}}. \added{While there are varying definitions of the mass transfer efficiency in the literature}, in this study we use the definition $\beta = \dot{M}_\mathrm{lost}/ \dot{M}_\mathrm{RLOF}$, the fraction of RLOF transferred mass that is lost from the system, \added{i.e. the transferred mass that is not accreted onto the accretor} \citep{tvdh06}. \replaced{fraction of mass not accreted ($\beta$)}{Mass} leaves the system with the specific orbital angular momentum of the accreting star, while ($1-\beta$) is \replaced{accreted onto}{accepted by} the accretor. \added{In our grid we use $\beta$ values} of $0.0$ \added{(conservative, no mass lost from the system)}, $0.25, 0.50$ and $0.75$ \added{($75\%$ of the mass leaves the system)}.

\subsection{Description of the Grid}\label{sect:grid}
\deleted{While the electron capture channel for single stars is fairly narrow \citep[cf.][]{phl+07,sie07} and measures about $0.25\msun$ around an initial mass of $9\msun$ (depending on the adopted treatment of convection this mass can shift, even though the width of the channel remains approximately the same) \citet{plp+04}, however, suggested that the mass range ECSN in binaries could  potentially be much wider, starting at a primary mass of $7\msun$ for wide binaries (Case B mass transfer)\footnote{We follow the definitions of ~\citet{tvdh06}: In Case A mass transfer the donor star fills its Roche-lobe during core-hydrogen burning. In Case B mass transfer the donor stars fills its Roche-lobe during hydrogen shell burning. In Case C mass transfer the donor fills its Roche-lobe during or after core-helium burning.} and ending at a primary mass of $17\msun$ for very close binaries (Case A mass transfer).} \replaced{To include all possible scenarios}{To cover the full range of possible mass transfer scenarios in binaries,} we calculated a dense grid of models with initial primary masses between $8.0\msun$ and $14.5\msun$ spaced by $0.25\msun$, mass ratios ($q=M_2/M_1$) between $0.6$ and $0.9$ with a spacing of $0.1$ and initial periods between $13$ and $35$ days with an interval of $1$ day. A second, denser, grid was calculated with initial primary masses between $13.3$ and $15.0\msun$ with a spacing of $0.1\msun$, mass ratios between $0.65$ and $0.95$ with a spacing of 0.05, and initial periods between $3$ and $12$ days with a interval of $1$ day ($0.5$ days below $4.0$).

Binary systems in \replaced{this parameter range}{in our grid}  generally undergo Case A or Case B mass transfer, with the primary losing a significant fraction of its mass\deleted{generally} before carbon burning commences. As we will discuss below, \replaced{low}{short} period and \replaced{high}{long} period systems (early Case A and Case B) and systems with extreme mass ratios are more prone to develop a contact system \citep[cf.][]{wlb01}. Once systems \replaced{develop}{enter into} contact, we don't follow their evolution \replaced{anymore}{further} and ignore them in our analysis.  \replaced{Since MESA does not have the capability to compute models through common envelope evolution and beyond (e.g. mergers),}{Since presently MESA does not offer robust methods to compute common envelope evolution and/or mergers, it is unpractical to include these evolutionary paths in a parameter exploration study requiring large numbers of models. Therefore,} we did not investigate models with initial periods below $3$ days as in \citet{me16}.
To avoid unnecessary computations in the late phases of the evolution we terminated the models at either neon ignition \added{(which is for primaries in this mass range below $\log (\rho_{c}) = 8.25$ g\,cm$^{-3}$, see Fig.~\ref{fig:tcrhoc_mloss-late})} or at $\log (\rho_{c}) > 8.5$ g \,cm$^{-3}$, whichever comes first. To establish the final evolution of stars that did not ignite neon, we computed several models beyond $\log (\rho_{c}) > 8.5$ g\,cm$^{-3}$ and results of these models will be discussed in Sec.~\ref{sect:new_ECSN_pathway}.

\subsection{Relevant Definitions for ECSN}
We assume that stars with carbon-oxygen core masses (hereafter $M_\mathrm{CO}$) below $1.37\msun$ will not develop conditions conducive for electron captures on neon and magnesium \citep{jhn+13, tyu13}, and that stars that develop neon burning (either central, or for this mass range more likely, off-center) will end their lives as a CCSN \citep[e.g.][]{jhn+13, wh15}. A recent study by \citet{tlp15} indeed found that helium stars in close binaries with $M_\mathrm{CO}$ in the $1.37 - 1.43\msun$ mass range exploded as ECSN. While the threshold for neon burning in single stars is found at approximately $M_\mathrm{CO}$ = $1.42\msun$ \citep[e.g.][]{jhn+13, wh15}, this is not necessarily the case for primaries in close binaries. As noted by \citet{tlp15}, in addition to a core mass in a critical range, the core also needs to have a sufficiently high temperature for the onset of neon burning.

For stars that develop CO cores with $M_\mathrm{CO}$ above $1.37\msun$ but do not ignite neon, we employ a method \added{similar to the one} first pioneered by \citet{nom84e} to determine the final fate based on the mass the core. \added{However, }since stars in our mass range undergo significant mass \replaced{loss}{transfer through RLOF}, which \replaced{also}{may} affect the mass of the helium core \citep{wl99b}, we \replaced{decided to}{instead} use the mass of the CO core instead of the helium core\deleted{(as in \citealt{nom84e})} to determine the final fate of these stars. Throughout this paper we use the following definitions for the various core masses. A helium core boundary is defined at the outermost location where the hydrogen mass fraction is below 0.01 (i.e. a hydrogen free core). Similar definitions are used for the CO core (outermost mass location where the helium mass fraction is below 0.01), and the ONe core (outermost mass location where the carbon mass fraction is below 0.01).
 \deleted{Based on single star models there is consensus in the literature that a star will experience an ECSN when $M_\mathrm{CO}$ is between $1.37$ and $1.42\msun$ \citep[see for example][who all adopt similar values]{wh15,tlp15,me16}. }

\added{In section~\ref{sec:new_ECSN_mass_range} we analyze the conditions in stars that do not develop neon ignition. To determine whether conditions at a certain location are conducive for neon burning we calculated for a composition of $50\%$ neon and $50\%$ oxygen the neon burning energy generation and neutrino loss rate for densities and temperatures representative for our cores. The energy generation rate for neon burning was taken from \citet{whw02} and the neutrino loss rate from \citet{ihh+96}, through Fortran routines provided by F.X. Timmes, available at \url{http://cococubed.asu.edu/code_pages/nuloss.shtml}. This allows us to infer, in terms of density and temperature, whether the conditions for dominant neon burning are met or not.}

\section{Representative Systems: Case A and B}\label{sec:rep_systems}

Before we discuss the details of our grid, we first consider a sample of representative binary evolution systems that illustrate the variety of outcomes that systems in this parameter range can experience. Every system is calculated with four different values of $\beta$ from $\beta = 0$ to $\beta = 0.75$. As argued by \citet{pac81}, a little amount of matter can spin up the accretor to critical rotation, and it is assumed that the star can not accrete any more matter. However, it is still unclear how the star regulates exactly how much it accretes \citep[cf.][]{wlb01, mli+13}, and there is evidence for both close-to-conservative systems \citep{lyp+03} and for systems with close-to-non-conservative evolution \citep{mph07}. Therefore, we take $\beta = 0.5$ as our reference case and discuss higher and lower mass transfer efficiencies for each case. This choice of $\beta$ might be too high still, unless a disk is able to mediate the simultaneous accretion of matter and loss of angular momentum from the star through viscous stresses \citep[cf.][]{pac91, dsd+13}.

\subsection{Case A Evolution}\label{sec:Case_A}
\replaced{Stars that start mass transfer during their core hydrogen burning stage are considered Case A mass transfer systems}{Systems where mass transfer starts during the core hydrogen burning phase of the donor are defined to be undergoing case A evolution.}\deleted{Generally mass transfer proceeds on the nuclear timescale, which leads to a moderate ($\sim 10^{-6} \msun$\,yr$^{-1}$) mass transfer rate. }\deleted{In order t}To start mass transfer during the core hydrogen burning stage of the primary star the system needs to have an orbital period below $3$ days\added{, although this is somewhat dependent on the initial mass and mass ratio \citep[c.f.][]{wlb01}}. Many \deleted{of the} systems in our grid with Case A mass transfer lead to contact and could produce an ECSN through the merger scenario seen in Fig. 1a,b of \citet{me16}, or avoid evolution into an ECSN altogether. However, systems with periods of approximately $3 - 3.5$ days, avoid contact and are able to produce a well developed CO core in the mass range for ECSN to occur. 

\begin{figure*}[tbp]
\begin{center}
\includegraphics[scale=0.58]{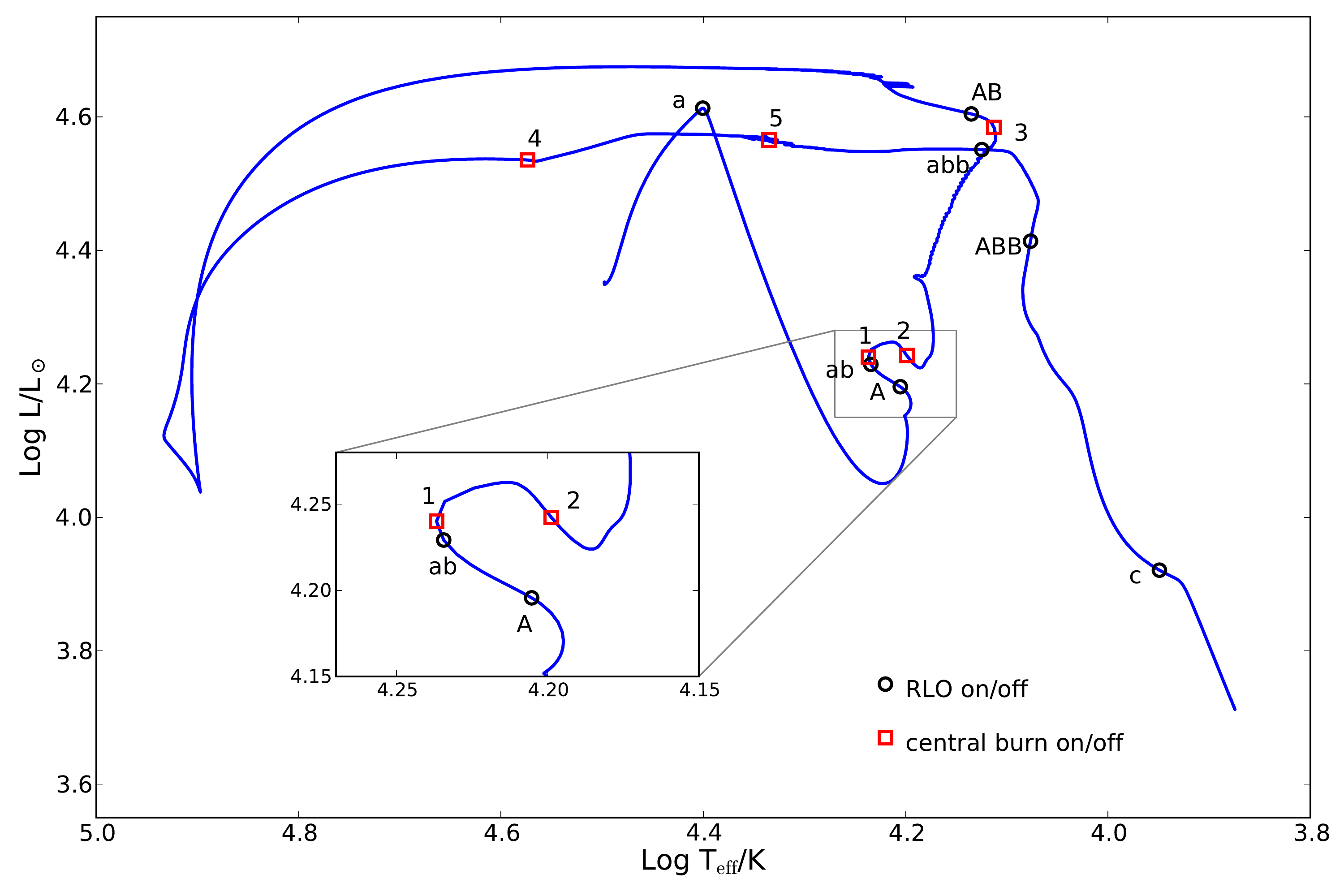}
\caption{HR diagram of the evolution of the primary star in a system with a primary mass of $15.7 \msun$, a secondary mass of $12.56\msun$ (\replaced{$q = M_2/M_1 = 0.8$}{$q = 0.8$}), an initial orbital period of $3$ days and a $\beta$ parameter of $0.5$. The start and end of RLOF is indicated with a black circle: case A (on:a, off:A), case AB (on:ab, off: AB), case ABB (on:abb, off:ABB), case C (on:c). The start and end of central nuclear burning is indicated with a red square: central hydrogen burning off (1), central helium burning on (2), off (3) and central carbon burning on (4) and off (5). \deleted{The primary star fills its Roche-lobe during the main sequence (black circle at $\log T/\mathrm{K} = 4.4$). It continues to lose mass until core hydrogen burning is almost finished, but it picks right up again (respectively black diamond, red square, and black circle in inset). RLOF starts again during the core helium burning phase (red triangle in inset) until it ceases right after central helium burning stops (red square and black diamond at $\log T/\mathrm{K} \approx 4.12$. During helium shell burning the star shrinks significantly, moving all the way to the blue (left) side of the HRD, after which eventually central carbon burning starts (red triangle at $\log T/\mathrm{K} = 4.58$. This ceases, and the star experiences several carbon flashes, during which the star expands and starts two more RLOF episodes. As a result of vigorous mass loss in the last phases of its evolution the star dims considerably by the end of the model run. The  model was terminated because of instabilities in the envelope.}}
\label{fig:CaseA_HRD}
\end{center}
\end{figure*}

\begin{figure}[t!bp]
\begin{center}
\includegraphics[scale=0.55]{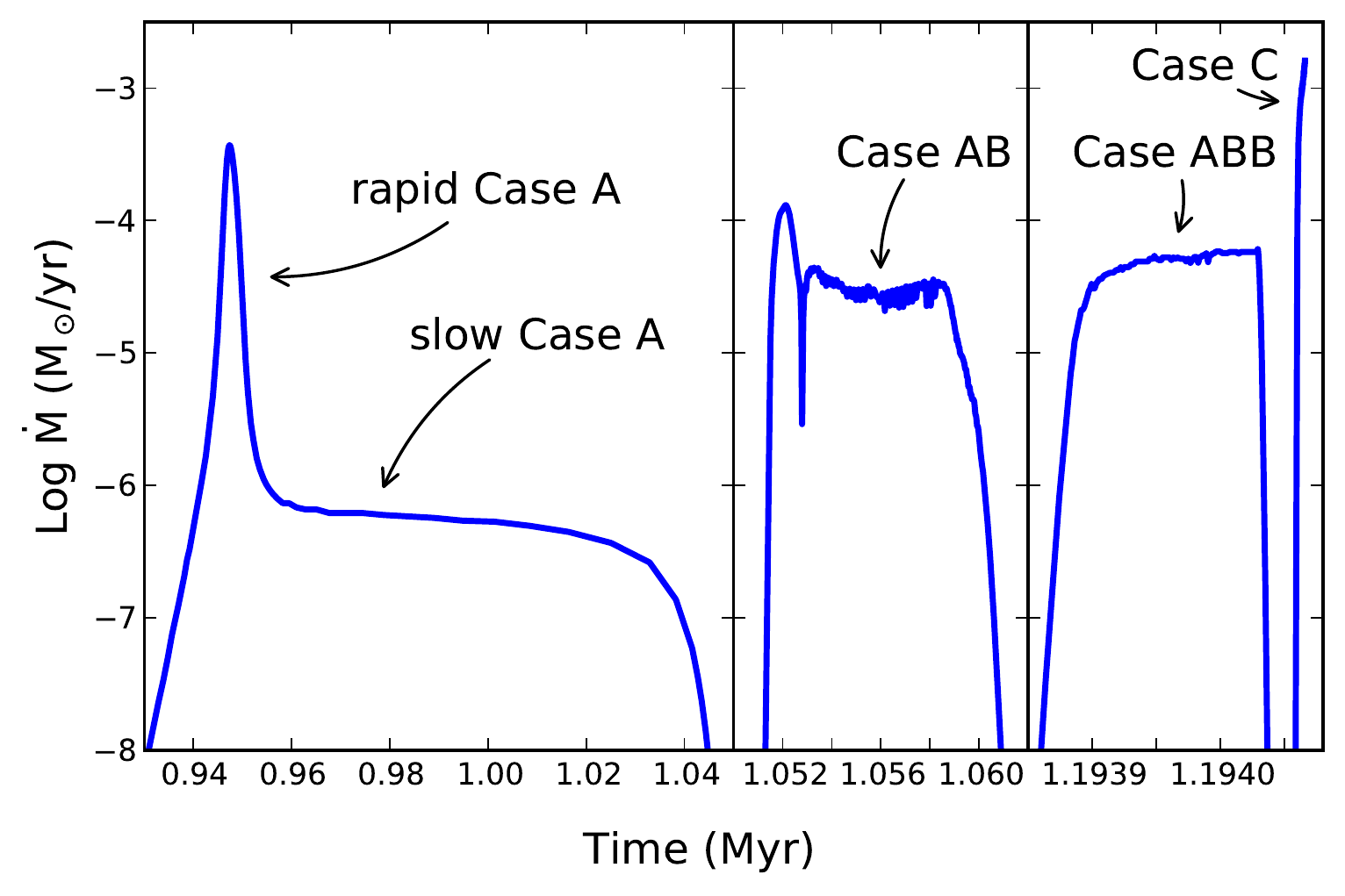}
\caption{Details of the mass transfer rate as a function of time for Case A system. Three distinct phases are visible. The left panel shows the mass transfer \replaced{advancement}{rate} during core hydrogen burning. The middle panel shows the mass transfer rates during hydrogen shell burning. The last panel shows mass transfer rates during and after carbon burning, cf. Fig.~\ref{fig:CaseA_Kipp}. The rise in mass transfer rate at the very end is a result of instabilities arising in the envelope and will be discussed in Sec.~\ref{sect:new_ECSN_pathway}.}
\label{fig:CaseA_mdot}
\end{center}
\end{figure}

\begin{figure}[tbp]
\begin{center}
\includegraphics[scale=0.57]{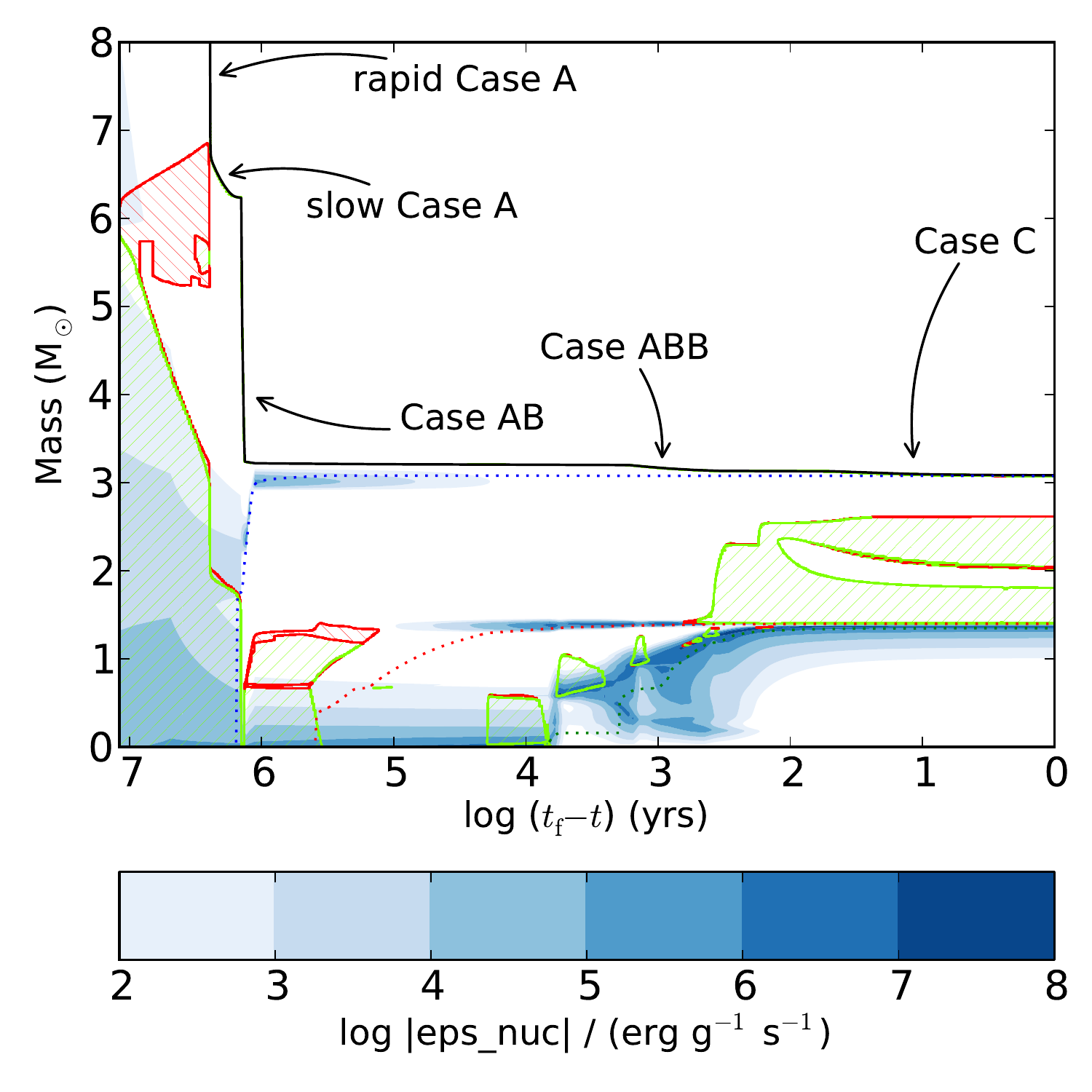}
\caption{\replaced{Evolution of the convective, mixing and burning regions in}{Kippenhahn plot of} the \replaced{primary component}{donor} of our Case A system.  The total mass is indicated by the solid black line. The green hatched regions indicates areas of convection, while the red hatched regions indicate semi-convection. Regions of nuclear energy generation are marked with blue shading. Only the inner $8 \msun$ of the \added{initially 15$\msun$} star are shown. \deleted{Mass transfer starts during the main sequence (Case A mass transfer, indicated with an arrow), and is followed by periods of mass transfer during hydrogen shell burning and carbon burning.} As the later phases in the evolution take place on a much shorter timescale, the x-axis is plotted with a logarithmic scale.}
\label{fig:CaseA_Kipp}
\end{center}
\end{figure}

\replaced{The evolution of one such system, representative for all systems in this class, is shown in Figs.~\ref{fig:CaseA_HRD} (Hertzsprung-Russell Diagram), ~\ref{fig:CaseA_mdot} (mass loss history), and \ref{fig:CaseA_Kipp} (Kippenhahn diagram)}{We show the evolution of a representative system of this class on the Hertzsprung-Russell diagram in Fig.~\ref{fig:CaseA_HRD}, its mass loss history in Fig.~\ref{fig:CaseA_mdot} and the evolution of its internal structure on the Kippenhahn diagram in Fig.~\ref{fig:CaseA_Kipp}}. This particular system has an initial primary mass of $15.7 \msun$, a secondary mass of $12.56\msun$ (\replaced{$q = M_2/M_1 = 0.8$}{corresponding to $q = 0.8$}), an initial orbital period of $3$ days and is evolved with $\beta = 0.5$.\deleted{(i.e. half of the transferred mass leaves the system through a fast wind; the other half is accreted onto the secondary)} The primary reaches its Roche lobe after $0.94 \times 10^7$ years (see Fig~\ref{fig:CaseA_HRD}, \ref{fig:CaseA_Kipp}) when the helium mass fraction in the core is $0.89$. \replaced{, which started mass transfers through Roche lobe overflow.}{This initiates mass transfer through RLOF.} As seen in Fig.~\ref{fig:CaseA_mdot}, this initially takes place at a fairly high rate (thermal time scale) \added{as the donor adjusts to the decreasing orbital separation}. This is the so-called rapid Case A mass transfer phase \citep{pol94, wlb01}, during which the primary loses about $9 \msun$. After the mass ratio reverses, the mass transfer slows down to values around $\sim 10^{-6}\msun$\,yr$^{-1}$, driven by the nuclear evolution of the star. 

During this phase the star loses another $0.4\msun$, and finishes this first mass transfer phase with a total mass of $6.23\msun$, and a helium core mass of $1.75\msun$. The ignition of the hydrogen burning shell causes the star to expand\added{, attempting to grow to red giant dimensions}, which starts the second mass transfer phase, the so-called Case AB phase \citep[\added{i.e. a case B mass transfer phase following a case A,}][]{wlb01}, which proceeds on the thermal timescale of the star. During this phase mass transfer rates are of the order of $3 \times 10^{-5}\msun$\,yr$^{-1}$, and the star loses an additional $3\msun$ so that the total mass left is $3.24\msun$ (which, in this system, corresponds to the mass of the convective hydrogen core at the time Case A mass transfer started, see Fig.~\ref{fig:CaseA_Kipp}). At this time the helium core measures $2.13 \msun$, which through subsequent hydrogen shell burning \replaced{growed}{grows} to $3.08 \msun$. Once Case AB mass transfer finishes, \added{the star is able to adjust its structure and thereby its radius to recover hydrostatic equilibrium. As a result, the radius of the star decreases, the temperature increases}\deleted{while the luminosity stayed relatively constant},\deleted{which resulted in a shrinking of the star in a} and the star moves to the left side of the Hertzsprung Russell diagram, becoming a hot and compact helium star (see Fig~\ref{fig:CaseA_HRD}). The star finishes helium core burning at $1.172 \times 10^7$ yr and \replaced{developed}{subsequent core compression ignites} helium shell burning\added{, which expands the star again to red giant dimension, starting another mass transfer episode (Case ABB).} By the time that convective carbon burning starts in the core, a CO core of $1.29\msun$ has formed. Throughout the carbon burning phase (various carbon flashes), which lasts from $1.192 \times 10^7$ yr till $1.194 \times 10^7$ yr (about $20,000$ yrs), the star forms a core composed of neon and oxygen. Another episode of mass transfer starts during the final carbon flashes (the third panel in Fig.~\ref{fig:CaseA_mdot}, also marked in Fig.~\ref{fig:CaseA_Kipp}) which erodes the final bit of the remaining hydrogen layer and cancels hydrogen burning. 

By the end of carbon burning the ONe core has a mass of $1.19\msun$ while the CO core has a mass of $1.34\msun$. During the last carbon shell flash a convective shell forms on top of the helium burning shell, which slowly diminishes the helium burning intensity. This also slows down the growth of the CO core, which is able to grow to a mass of $1.40\msun$ by the end of the model run. Significant mass loss develops at the very end of the model run and causes instabilities that terminate the model (see also Section~\ref{sect:new_ECSN_pathway}). During the evolution of the primary, the secondary has grown by $\sim 6\msun$ to a total mass of $18.58\msun$. Although it is now much more massive than the primary, it is still a main sequence star (hydrogen mass fraction of $0.03$) due to the accretion of large amounts of fresh hydrogen, causing the star to rejuvenate \citep{hel83, msl+14,spl+16}. If the primary is able explode and forms a neutron star, this system \replaced{might}{most likely} evolves into an Be/X-ray binary.

\added{Systems with a different mass transfer efficiency display a similar behavior, although the orbit widens more rapidly for less conservative mass transfer, and less rapidly for more conservative mass transfer. This results in slightly larger final $M_\mathrm{CO}$ for stars with $\beta = 0.75$ and slightly smaller final $M_\mathrm{CO}$ for stars with $\beta = 0.0$ and $0.25$, compared with our fiducial case of $\beta = 0.5$. Due to the different rates at which the orbit responds to mass transfer and loss for various values of $\beta$, the number and the intensity of RLOF episodes also shows some variation, but since mass transfer is primarily driven by changes in the thermal structure of the primary, in the end, similar final outcomes are found.}

\subsection{Case B Evolution}\label{sec:Case_B}
\begin{figure*}[htbp]
\begin{center}
\includegraphics[scale=0.58]{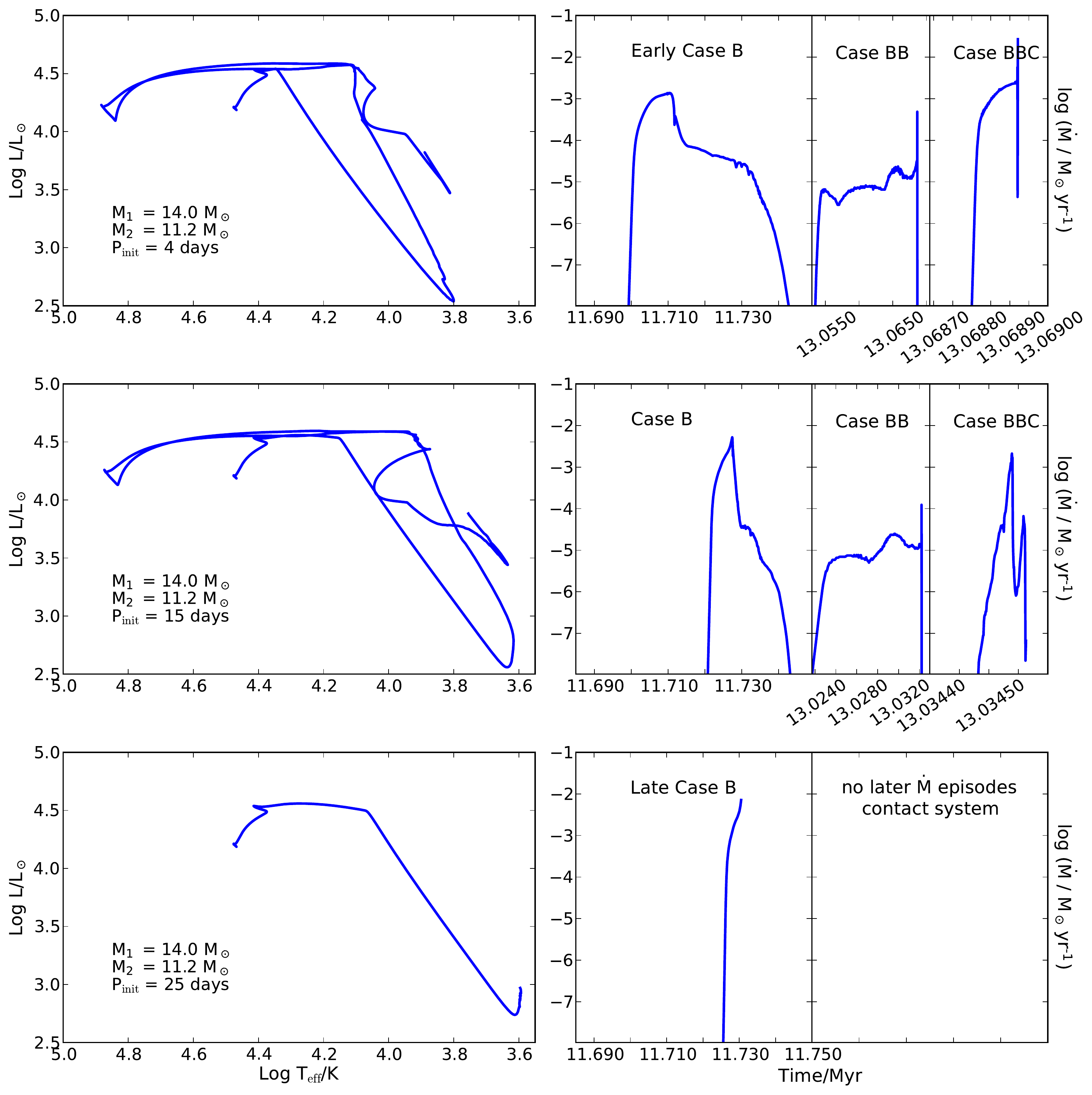}
\caption{Evolution of \replaced{three different systems}{a binary system with $M_1 = 14.0 \msun$ and $M_2 = 11.2 \msun$, with three different initial orbital periods, corresponding to early Case B mass transfer (4 days, top row), intermediate Case B mass transfer (15 days, middle panel) and late Case B mass transfer (25 days, bottom panel)}.\deleted{The blue line indicates a system with an initial orbital period of $4$ days. The green line corresponds to $15$ days, and the red line to $25$ days. Top Left: The evolution of the systems in a Hertzsprung-Russell diagram. Top Right: The evolution of the mass transfer rates. The red line ($P = 4$ days) represents a model that develops into contact. Bottom Left: The evolution of the stellar (solid) and Roche radii (dashed). Bottom Right: The evolution of the orbit. The systems with the smallest periods ($4$ and $15$ days respectively) evolved into a relatively wide binary (after several mass transfer episodes), while the system with initial period of $25$ days developed high mass transfer rates and subsequently contact during the first mass transfer phase.} The left column shows the Hertzsprung-Russell diagram for the three systems, the right column shows the mass transfer history as a function of time for the three systems. While the first window of each mass loss panel has the same scale and allows for comparison between the timing of the onset of mass transfer, later phases are zoomed in considerably to show detail and have different scales on the time axis.}
\label{fig:CaseB-typical_combined}
\end{center}
\end{figure*}

Stars that \replaced{start}{undergo their first} mass transfer \added{phase} during hydrogen shell burning are considered Case B mass transfer systems. \replaced{Generally,}{The removal of high-entropy layers of the envelope causes he star to shrink on a dynamical timescale, however, as the donor expands on the thermal timescale during the crossing of the Hertzsprung gap this leads to stable but high}  mass transfer rates of $\sim 10^{-4} -10^{-3} \msun$\,yr$^{-1}$\added{, with a distinguishable fast and subsequent slower rate as the orbit initially shrinks and then widens again (c.f., Fig~\ref{fig:CaseB-typical_combined})}. \added{Differences in the timing of the start of mass transfer, due to different initial periods, lead to significant different evolutionary paths.} Early Case B mass transfer will \added{appreciably} affect the formation of the helium core as the intensity of hydrogen shell burning is diminished as a result of the envelope responding to the high mass loss rates. Late Case B mass transfer will have less of an effect as the helium core has already been established by the time mass transfer starts. \added{Variations in the value of $\beta$ also cause notable differences. Systems that evolve conservatively or nearly conservatively, have a higher chance to develop contact, because of two associated effects. As less mass is lost from the system, the accretor gains mass faster, and hence fills its Roche lobe earlier, leading to more contact systems. This effect is stronger for systems with $q<<1$ as the initial Roche lobe of the accretor is much smaller than for systems with a mass ratio close to $1$. However, in addition, initially wider systems, develop higher mass loss rates as they are further into the Hertzsprung gap \citep[c.f.][]{wlb01, pol94} which pushes the accretor further out of equilibrium. As the accretor is still a main sequence star and has a radiative envelope, mass accretion makes it swell considerably, leading to contact. On the other hand, systems that evolve more non-conservative than our fiducial model show more contact free evolution up to higher initial periods, because the accretor does not gain as much mass and accretes it at lower rates  as  most of the mass is leaving the system, making it less prone to swell up and reach its Roche lobe.} \replaced{As the timing of the start of mass transfer will affect the subsequent evolution of the system we will}{Because the time of the beginning of Case B mass transfer has a strong influence on the subsequent evolution, we} discuss three different Case B systems in order to establish a clear picture of Case B evolution.

Fig.~\ref{fig:CaseB-typical_combined} shows the evolution of three Case B systems (an early, mid, and late Case B system). System 1 (top row), an early Case B system ($M_1 = 14.0\msun$, $M_2 = 11.2\msun$,  $P_\mathrm{init} =   4$ days), started mass transfer promptly after the primary star finished core hydrogen burning and started crossing the Hertzsprung gap ($t_\mathrm{RLOF1}$ = 11.6998 Myr). Mass transfer rates of $\sim 10^{-3}\msun$\,yr$^{-1}$ are seen for approximately $10,000$ yrs during which \added{the orbit shrinks and} the primary loses $9.8\msun$. After the \replaced{initial fast Case B phase,}{mass ratio reverses and the orbit widens again as a response to the mass transfer, the star is able to adjust its thermal structure, and around $11.71$ Myr} a period of slower mass transfer \replaced{continued}{commences (slow Case B, \citealt{pac71, doo84})} with rates of $\sim 3 \times 10^{-5}\msun$\,yr$^{-1}$ (Fig~\ref{fig:CaseB-typical_combined}, top right). The separation of the binary components increases from $31\rsun$ \replaced{before}{at the start of} Roche lobe overflow to $\sim 100\rsun$ after the fast Case B phase, and $150\rsun$ ($48$ days) after the first mass transfer phase finishes (i.e. when helium ignited in the core). \added{During hydrogen shell burning, the core of the star continues to contract until helium ignites in its core. This allows the star to readjust its thermal structure again, and }the star shrinks in response \replaced{helium core burning}{to this, }\added{detaches from its Roche lobe,} and experiences a pause in mass transfer. \replaced{This allows the star}{As a result, the star is able} to evolve through core helium burning relatively unaffected. After helium core burning the primary has a mass of $3.3\msun$, with a helium core of $3.11\msun$ and a CO core of $1.29\msun$. \added{Continued contraction heats up the core and layers above, until the helium shell ignites. The envelope readjusted its thermal structure once again through expansion, fills its Roche lobe and restarts} mass transfer through Roche lobe overflow \deleted{picked up again} during the \replaced{major phase of}{subsequent} core carbon burning \added{phase (case BB, $t_\mathrm{RLOF2}$ = 13.0534 Myr)}. \added{This phase only lasts} \deleted{and lasts} for $\sim 20,000$ years at a rate of $\sim 5 \times 10^{-5}\msun$\,yr$^{-1}$ during which the star loses its residual hydrogen layer of $0.2\msun$. \added{A final mass loss episode starts at $t_\mathrm{RLOF3} = 13.0688$ Myr (Case BBC), which after $\sim 100$ years leads to instabilities that terminate the model.} At the end of its evolution the star has established an helium core of $3.\replaced{117}{12}\msun$, a CO core of $1.\replaced{399}{40}\msun$ and an ONe core of $1.\replaced{349}{35}\msun$. The final orbital separation of the system is $\sim 175\rsun$. 

\added{The same system, evolved with different values of $\beta$ gives fairly similar evolutionary outcomes, although with slightly less massive CO cores for more conservative evolution and slightly more massive CO cores for less conservative evolution. All systems avoid contact, due to their small initial orbital period, but the systems with more conservative mass transfer develop smaller CO cores as a result of a tighter system and hence prolonged mass transfer.}

\begin{figure}[htbp]
\begin{center}
\includegraphics[scale=0.57]{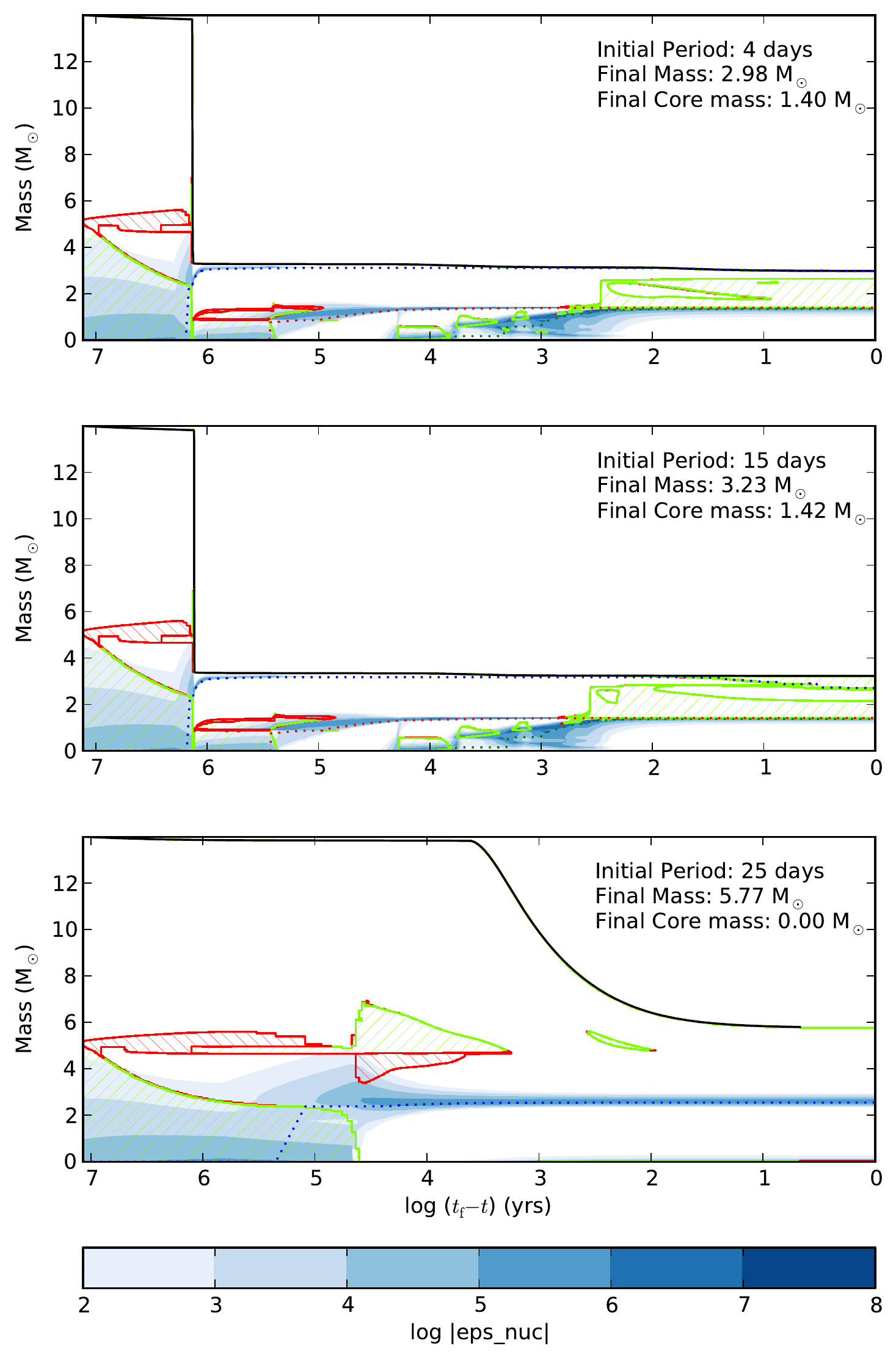}
\caption{Kippenhahn diagrams for three Case B systems discussed in the text.\deleted{The upper panel corresponds to the system with an initial period of $4$ days, the middle panel corresponds to the system with an initial period of $15$ days, and the bottom panel corresponds to a system with an initial period of $25$ days.} The total mass is indicated by the solid black line. The green hatched regions indicates areas of convection, while the red hatched regions indicate semi-convection. Regions of nuclear energy generation are marked with blue shading. Time is plotted on the $x$-axis, with the value of $t_\mathrm{f} - t$/yr indicating the remaining time in the calculation, plotted logarithmically to show detail in the later phases of the evolution.\deleted{ Taking the logarithm of this quantity ensures that the later phases in the evolution, which take place on a much shorter timescale than the main sequence, are clearly distinguished. As the value for $t_\mathrm{f}$ is different for each of the models, a direct comparison between positions on the $x$-axis among the three models cannot be made.}}
\label{fig:CaseB_Kip-comb}
\end{center}
\end{figure}

System 2 (Fig.~\ref{fig:CaseB-typical_combined}, \replaced{green line}{middle row}) is a binary system ($M_1 = 14.0\msun$, $M_2 = 11.2\msun$, $P_\mathrm{init} =  15$ days) that experiences the start of Roche lobe overflow in the middle of the Hertzsprung gap ($t_\mathrm{RLOF1}$ = 11.7210 Myr). \added{As a result of the natural expansion of the star and the shrinking orbit,} mass transfer rates quickly increase to values of $\sim 5 \times 10^{-3}\msun$\,yr$^{-1}$ during which $10.4\msun$ is lost from the primary, leaving a donor with a mass of $3.6\msun$, a helium core of $3.18\msun$ and a CO core of $1.31\msun$ at the completion of core helium burning. \replaced{After a short break}{Due to adjustments to the thermal structure of the star during the contraction of the CO core}, mass loss increases again during the carbon burning phase (Fig.~\ref{fig:CaseB_Kip-comb}, middle row, Case BB, $t_\mathrm{RLOF2}$ = 13.0237 Myr), but now at relatively low rates of $\sim 7 \times 10^{-5}\msun$\,yr$^{-1}$. This moderate mass loss removed almost the entire residual hydrogen envelope, and leaves a star with mass of $\replaced{3.230}{3.23}\msun$, a He core of $\replaced{3.223}{3.22}\msun$, a CO core of $\replaced{1.417}{1.42}\msun$ and a ONe core of $\replaced{1.381}{1.38}\msun$\deleted{ is left}. \added{A third, case BBC, mass transfer phase, starts at $t_\mathrm{RLOF3}$ = 13.0345 Myr, but only removes $0.01 \msun$ before it terminates due to instabilities that develop in the envelope.} During the fast Case B mass transfer period, the orbit increased from $75\rsun$ (initial) to $345\rsun$, which increases during subsequent mass loss episodes to $370 \rsun$. While the initial period of this system was quite a lot larger than system 1, the evolution of system 2 proceeded very similar to that of system 1. While system 1 starts mass transfer earlier, system 2 reaches higher mass transfer rates due to its higher mass and shorter Kelvin-Helmholz timescale. Both mass loss episodes terminate when helium ignites in the core (roughly about the same time, $t = 1.1742 \times 10^{7}$ yr, see Fig.~\ref{fig:CaseB-typical_combined}), and consequently the amount of mass transferred during the first mass transfer phase is about equal ($\sim 10\msun$).

\added{Evolving this system with different values of $\beta$ shows that the most conservative scenarios ($\beta = 0, 0.25$) lead to contact during RLOF, while the systems with $\beta = 0.5, 0.75$ are able to avoid contact. This is the result of a faster growing, and thus quicker swelling accretor in the case of conservative mass transfer, while systems with largely non-conservative evolution don't accrete as much matter onto the secondary, avoiding an expansion of their radius beyond their Roche lobe and hence contact all together.}

System 3 (Fig.~\ref{fig:CaseB-typical_combined}, bottom row) shows the evolution of a binary system ($M_1 = 14.0\msun$, $M_2 = 11.2\msun$, $P_\mathrm{init} =  25$ days) that experiences the start of Roche lobe overflow late in the Hertzsprung gap ($t_\mathrm{RLOF1}$ = 11.7250 Myr). As the primary expands beyond its Roche lobe, mass transfer rates quickly ramp up to values of $7.7\times 10^{-3}\msun$\,yr$^{-1}$. Although the orbital separation increases, and thus the size of both Roche lobes, the secondary quickly fills its Roche lobe due to the large amount of accreted matter, leading to a contact system after only $4500$ yrs \citep[cf.][]{mpy08, lan12, mli+13}.

When evolved with different values of $\beta$, only the $\beta = 0.75$ is able to avoid contact, as the lower mass accretion rates and faster growing orbit allow the accretor to respond more efficiently to mass accretion, leading to less swelling and avoiding contact.

\section{The Effects of Mass Loss on the Evolution of the Primary}\label{sec:new_ECSN_mass_range}
In order to explode as an ECSN several ingredients need to be in place. \citet{nom84e} argued that stars with helium cores between $2.0$ and $2.5\msun$ (which corresponds to roughly to initial masses between $8$ and $10\msun$) would explode as an ECSN. His models, however, did not develop a second dredge-up which \added{can} significantly \replaced{reduced}{reduce} the mass of the helium core and \replaced{renders this criterion somewhat useless}{diminishes the predictive power of this criterion} \citep[cf.][]{plp+04,phl+07}. Since then, several authors \citep{sie07, phl+07, jhn+13, dsl+10, dgs+15} have established precise initial mass ranges for ECSN to occur, although these mass ranges are highly sensitive to the adopted convection criteria, overshooting and mass loss prescriptions \citep{phl+07, dsl+10, lan12}. \citet{jhn+13} produced several detailed models, computed all the way to electron captures on $^{24}$Mg and $^{20}$Ne, and found that CO cores \added{with masses} over \replaced{$1.35086\msun$}{$1.35\msun$} are able to reach densities high enough for this to occur ($\log \rho \approx 9.6$ g cm$^{-3}$). If  the CO core is massive enough, neon will ignite off center \citep{jhn+13, sqk16}, but \citet{jhn14} also found that the upper boundary for ECSN is affected by uncertainties regarding the progression or stalling of the neon flame. There seems to be consensus, however, that the mass of the CO core is a\deleted{fairly} reliable indicator for the final fate of stars in this mass range. \added{How this translates into the initial mass of the star, depends on the adopted convection criterion, with the Schwarzschild criterion producing more massive cores than the Ledoux criterion for the same initial mass. Inclusion of overshooting will also lead to larger cores \citep[cf.][]{sie07}. However, although the initial mass range for ECSN is therefore quite sensitive to the adopted convection criteria, this does not seem to be the case for the final $M_\mathrm{CO}$ as most authors find similar values for $M_\mathrm{CO}$ at which neon ignites, even though they treat convection differently. In the context of binary evolution, this provides an additional reason to adopt $M_\mathrm{CO}$ as indicator for whether the star explodes as an ECSN or not, as the CO core is generally not eroded by mass transfer, in contrast to the He core.}

The question, however, is whether the range in $M_\mathrm{CO}$ for ECSN in binaries is identical to the established $M_\mathrm{CO}$ range for ECSN in single stars. Our models indicate that this might not necessarily be the case. 
\deleted{To assess whether the primary star in a particular binary system evolves into an ECSN we likewise apply a criterion based on the mass of the CO core, rather than a criterion based on the mass of the helium core \citep{nom84e, plp+04}. In single stars the mass of the helium core is affected by the occurrence of the second dredge-up. Likewise in binary stars the mass of the helium core is affected by the onset and severity of mass loss. Generally, most of the hydrogen envelop is lost during mass transfer as a result of Roche Lobe Overflow \citep{tvdh06}, and thus the mass of the CO core at the completion of carbon burning is a more reliable indicator for the final fate than the mass of the helium core at whatever evolutionary stage. A recent study by \citet{tlp15} found that helium stars in close binaries with CO cores in the $1.37 - 1.43\msun$ mass range exploded as ECSN. This range is consistent with the values found by \citet{wh15} while the lower critical value is consistent with \citet{jhn+13} and \citet{tyu13}. However, as noted by \citet{tlp15}, in addition to a core mass in a critical range, the core also needs to have a sufficiently high temperature for both the onset of electron captures and neon burning respectively.}
Whereas CO cores with a mass of $\sim 1.42\msun$ in single stars would ignite neon in their cores \citep[cf.][]{nom84e,nom87c,jhn+13} our models do \replaced{show}{not show neon ignition due to} a slightly lower central temperature\deleted{ and therefore avoid neon ignition}. This rather different behavior is a direct result of the binary interaction, and thus unique to binary systems. We have identified, and will describe below, two processes that are responsible for that. \replaced{firstly,}{Compared to single stars,} \deleted{any} significant mass loss \replaced{leading up to}{before} the establishment of the CO core will \replaced{effectively reduce the}{create a smaller} CO core \deleted{mass} and \replaced{therefore the its}{lead to a consistently lower} temperature \added{for the rest of its life}. \replaced{ and, secondly,}{In addition,} mass loss during and after carbon burning will cause the \added{primary} star to cool down \replaced{much more}{faster} \replaced{than}{compared to} its single star counterpart\deleted{s} even if the core mass is the same.

\begin{figure}[tbp]
\begin{center}
\includegraphics[scale=0.68]{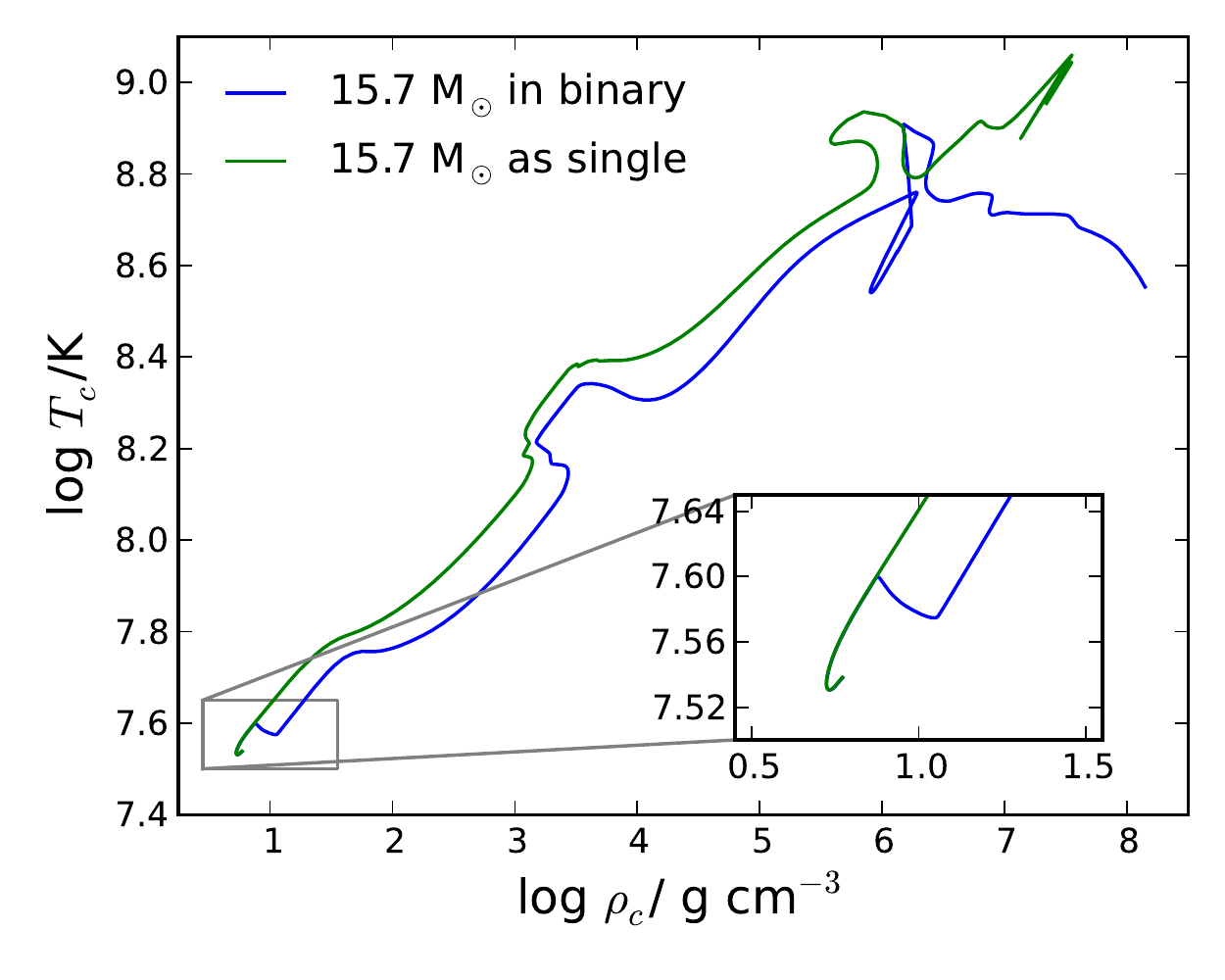}
\caption{Evolution of the central density versus central temperature. Shown in blue is our Case A model which undergoes mass transfer once the \added{core} temperature reaches $\log T/K = 7.6$. Shown in green is a single star model of the same initial mass ($15.7\msun$). \added{As shown in the inset,} the star that undergoes mass \replaced{loss}{transfer} develops a core that is cooler and more compact (higher density) \replaced{than the single star}{compared to a single star of the same initial mass}. The single star eventually develops off center neon burning and a CO core with a mass of $1.63 \msun$, while the star in the binary avoids neon burning and develops a CO core of $1.40\msun$.}
\label{fig:tcrhoc_comb_Case_A}
\end{center}
\end{figure}

\subsection{Mass Loss During Hydrogen or Helium Burning}
Significant mass transfer due to RLOF anytime before the establishment of the CO core will affect the \replaced{size}{subsequent formation} of the CO core. \added{In Fig.~\ref{fig:tcrhoc_comb_Case_A} we show the evolution of the core in the $\rho_\mathrm{c}$-$T_\mathrm{c}$ plane for the Case A system which we discussed in Sec.~\ref{sec:Case_A}. The evolution for a single star, which is not affected by mass transfer, is shown as a comparison. As a response to the high mass loss rates and to compensate for the loss of $9 \msun$ of matter during the fast Case A phase, the core of the primary tries to restore hydrostatic equilibrium by contracting to higher densities. This results in a convective core which is $1 \msun$ smaller than an undisturbed core (see Fig.~\ref{fig:CaseA_Kipp}). A new equilibrium is found at a central temperature which is 2.6 million Kelvin lower than the central temperature before mass loss.} \deleted{The star never recovers from this, and continues its evolution as if it were a less massive star.} \added{As a result, stars}\deleted{Stars} that \replaced{have undergone}{undergo Case A} mass transfer \deleted{as a result of RLOF} form smaller \added{convective} cores than their single star counterparts\added{, which results in the formation of smaller CO cores after He burning}. \deleted{If mass loss starts during the main sequence, it affects the mass of the convective core, resulting in a star that is effectively a lower mass star.} \deleted{For the Case A system that we discussed earlier, this can be seen in Fig.~\ref{fig:CaseA_Kipp} which shows the reduction of the convective core by $\sim 1\msun$. Figure~\ref{fig:tcrhoc_comb_Case_A} shows that, in order for the star to adjust to the high mass loss rates and to compensate for the loss of matter, the central temperature drops by $0.03$ dex. The star never recovers from this, and continues its evolution as if it were a less massive star. }

Case B systems do not show the notable decrease in central temperature upon the start of mass transfer, \added{as the hydrogen burning shell acts as a barrier between the exterior and the interior. As the core is contracting and heating up to allow for the start of He burning, the hydrogen shell is supporting the entire envelope independently of the core. However, as the envelope is quickly being reduced in mass, the outward progression of the hydrogen shell is limited by whatever is left over of the envelope, and hence not able to proceed as far outward as in cores not affected by RLOF (cf. Fig.~\ref{fig:CaseB_Kip-comb}).} \deleted{however, their cores are affected by mass loss. As Fig.~\ref{fig:CaseB_Kip-comb} shows, hydrogen shell burning is generally not able to proceed as far outward as a result of the large mass loss rates, which in turn leads to a smaller helium core. }

The net effect \added{of mass transfer before the establishment of the CO core} is that \added{the resulting} CO cores \deleted{of stars affected by mass loss during hydrogen and helium core burning are generally} are smaller \added{and cooler} than those of single stars. \deleted{This affects the subsequent evolution, as their internal structure is similar to that of CO cores with the same mass (e.g. originating from less massive single stars).}

\subsection{Mass Loss After Core Carbon Burning}\label{sec:late_mass_loss}

After central carbon burning, in the absence of a heat source, the core contracts again to maintain hydrostatic equilibrium. Due to this  contraction at high central densities, \added{thermal} neutrino losses increase and degeneracy sets in, resulting in a degenerate core which is governed by the balance between heating due to contraction and cooling due to neutrino losses \citep{pac71, nom84e, bbs+16}. In single stars, \added{hydrogen and} helium shell burning would cause the core to grow, and cooling as a result of neutrino losses would be offset by additional heating as a result of accretion due to core growth. \added{Even though a strong stellar wind, driven by helium shell flashes, can develop in the final phases of the lives of single stars, the center of these stars is able to maintain a more or less constant temperature, while contracting to densities conducive for electron captures to start.} However, all of our models experience significant mass \replaced{loss}{transfer} \added{due to RLOF during the carbon shell flash phase} (with rates up to, and sometimes exceeding, $\dot{M} > 10^{-3}\msun$\,yr$^{-1}$). This pushes the star out of thermal equilibrium, leading to additional cooling in the core as a result of an endothermic expansion to make up for lost envelope matter \citep[an effect similar to what has been \replaced{observed}{found} by][]{tlp15, sqk16}. \deleted{Because of the drop in temperature all throughout the star, it also reduces the intensity of helium shell burning, causing the core growth rate to decrease which subsequently accelerates the decrease in temperature. The net result is a rapidly cooling core, which never develops conditions for neon to ignite (only at a much higher $M_\mathrm{CO}$ of $\sim 1.52 \msun$) and never contracts to densities high enough for the electron captures on neon and magnesium to set in. }

\begin{figure*}[tbp]
\begin{center}
\includegraphics[scale=0.57]{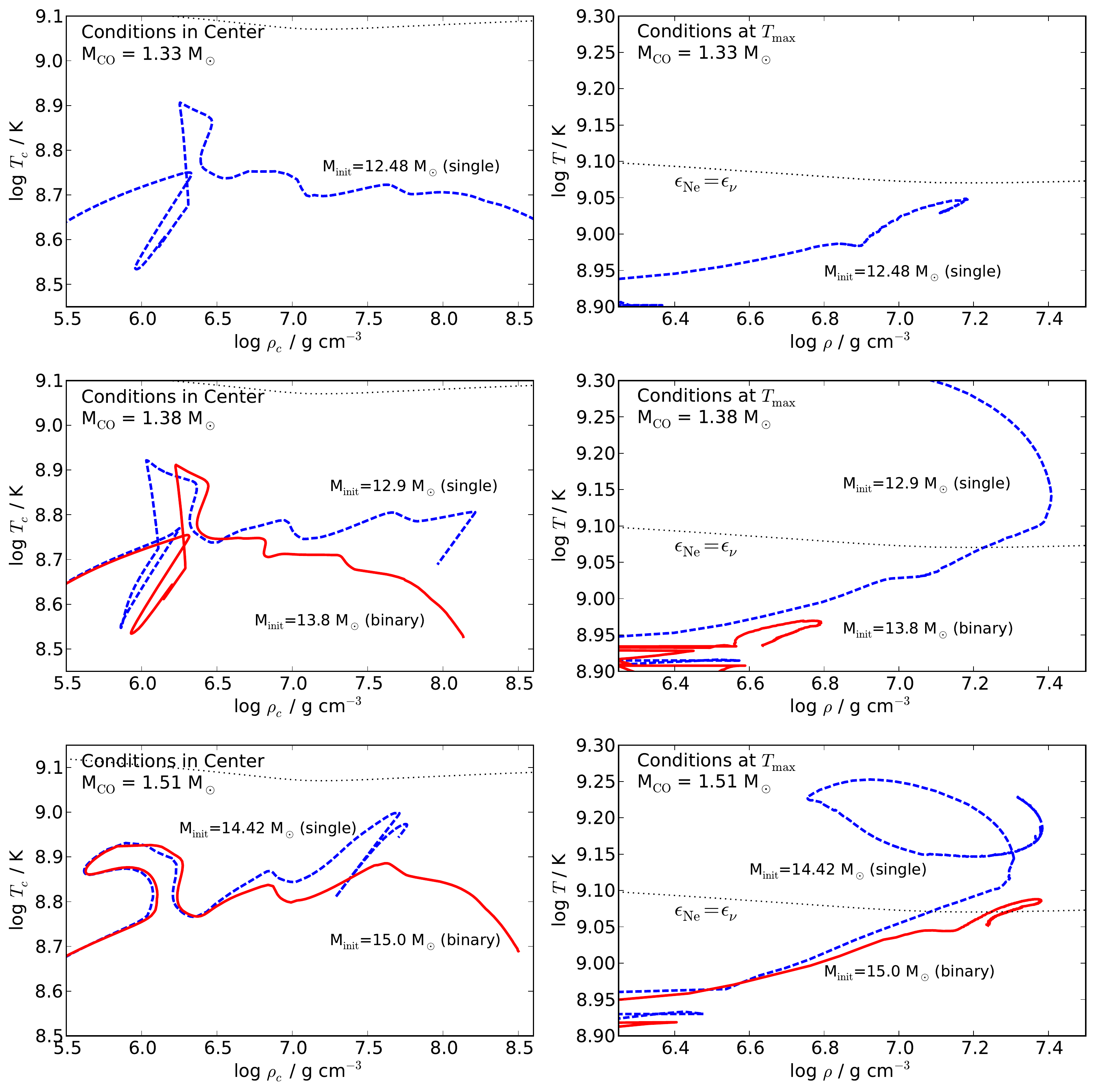}
\deleted{\caption{Evolution of the central density and central temperature of a single star model and the primary in a binary. The left panel shows in red a binary model with initial conditions ($M_1 = 13.8\msun$, $M_2 = 11.4\msun $, $P_\mathrm{init}=9$ days) that lead to a final $M_\mathrm{CO}$ of $1.391 \msun$. The red lines show a single star model with an initial mass of $12.9\msun$ that leads to a final $M_\mathrm{CO}$ of $1.389\msun$. The right panel shows a similar situation, but for higher mass models. The blue line shows a single star model with an initial mass of $14.42 \msun$ while the red line shows the primary stars in a binary system with initial conditions equal to $M_1 = 15.0\msun$, $M_2 = 12.0\msun $, $P_\mathrm{init}=9$ days.  Both the single star and the binary component have an $M_\mathrm{CO}$ of $1.40\msun$ prior to carbon burning and a final CO core mass of $1.51\msun$. Although the core masses are virtually the same for the single star and binary component through his phase of their evolution,  the single star develops neon burning (the hook and decrease in central density and temperature indicate an off-center flame), but the binary component does not. The core of binary component cools down while the core of the single star heats up. }}
\added{\caption{Evolution of the central density vs central temperature (left column) and density and temperature at the location of the maximum temperature (right column) for various single and binary models. The dotted line around $\log T/\mathrm{K} = 9.1$ indicates conditions where neon burning and neutrino losses are in balance. The top row shows the evolution of a single star with an initial mass of $12.48 \msun$. The middle row shows the evolution of a single star with an initial mass of $12.9 \msun$ and a binary component with an initial mass of $13.8 \msun$. Both develop a CO core with a mass of $1.38 \msun$, with the single star evolving through neon burning and the binary component avoiding neon burning. The bottom row shows the evolution of a single star with an initial mass of $14.42 \msun$ and a binary component with an initial mass of $15.0 \msun$. Both develop an CO core with a mass of $1.51 \msun$, with the binary component narrowly avoiding sustained neon burning.}}
\label{fig:tcrhoc_mloss-late}
\end{center}
\end{figure*}

\added{Fig.~\ref{fig:tcrhoc_mloss-late}} shows the\deleted{dramatic} effect of \deleted{the}high mass loss rates this late in the evolution of the star on\deleted{both} the central \deleted{and maximum}temperature. \added{The top row shows the central and conditions at $T_\mathrm{max}$ for a single star --not affected by mass transfer-- with mass of $12.48 \msun$, which has developed a CO core with a mass of $1.33 \msun$. After central carbon burning, the ONe core contracts to higher densities. Due to the weak temperature dependence of the electron degenerate pressure and the strong sensitivity of the plasma-neutrino emission rate to the temperature and density, a temperature inversion will develop in the ONe core \citep{nom84e}. Because of high density, the center of the star experiences strong neutrino cooling, while the layers further outward experience less cooling or even some heating due to their proximity to the carbon and helium shell sources. Eventually the location of $T_\mathrm{max}$ can be found somewhere between the center and the carbon burning shell \citep[cf. Figs. C1, C2 in][]{sqk16}. The conditions in the center are shown in the top left panel, while the conditions at $T_\mathrm{max}$ are shown in the top right panel. In single stars, where mass loss is due to stellar winds instead of RLOF, we find a steady increase of the maximum temperature as a result of core growth due to helium and carbon burning. In this case, the CO core is not massive enough to reach neon ignition, and as can be seen in the top right panel, the temperature reaches a maximum and then decreases again, showing a characteristic hook. This hook can be understood as the whole core now succumbing to intense cooling as a result of thermal neutrino losses, and the mass coordinate of $T_\mathrm{max}$ moving outward to lower densities, eventually merging with the carbon burning shell. If the core, however, is more massive (see middle and bottom panel of Fig.~\ref{fig:tcrhoc_mloss-late}), the conditions at $T_\mathrm{max}$ become conducive for neon ignition (shown by the blue dashed line crossing the $\epsilon_\mathrm{Ne} = \epsilon_\mathrm{\nu}$ line), causing the core to expand (shown by the hook in the blue dashed line in the left middle and left bottom panels). In the case of a CO core with a mass of $1.38 \msun$ ignition takes place at a radial mass coordinate of $0.95 \msun$, with ignition closer to the center for higher $M_\mathrm{CO}$ \citep{sqk16}.}

\added{When we compare, however, the evolution of the primary star in a binary (affected by significant past and ongoing mass loss) with the evolution of a single star as described above, we find some remarkable differences. The red lines in the middle and bottom row of Fig.~\ref{fig:tcrhoc_mloss-late} are for a binary component with the same $M_\mathrm{CO}$ as the single star shown in these panels. 
As can be seen in the middle left panel ($M_\mathrm{CO} = 1.38 \msun$ for both models), carbon burning in the binary component ($M_\mathrm{init} = 13.8 \msun$, $P_\mathrm{init} = 9$ days) takes place at a slightly higher density than in the single star, a consequence of the compacter core that has developed as a result of prior Case B RLOF. During the contraction phase after core carbon burning, however, the core cools down whereas the core of the single star heats up. The evolution of $T_\mathrm{max}$ is also different, with the single star rapidly evolving toward conditions conducive for neon ignition, but the binary component reaching a maximum temperature of $\log (T / \mathrm{K}) =  8.97$, by far not sufficient for neon ignition. The situation is similar for the binary model in the lower panel ($M_\mathrm{init} = 15.0 \msun$, $P_\mathrm{init} = 9$ days), which develops a CO core with a mass of $M_\mathrm{CO} = 1.51 \msun$. This is far above the accepted $M_\mathrm{CO}$ for neon ignition in single stars, but our model fails to produce sustained neon burning, even though it briefly develops conditions conducive for neon ignition.}

\deleted{ is shown in Fig.~\ref{fig:tcrhoc_mloss-late}. Because of prior neutrino cooling, a temperature inversion has developed in the interior of the star, with the location of the maximum temperature somewhere between the center and the carbon burning shell \citep[cf. Figs. C1, C2 in][]{sqk16}. In single stars, \replaced{(mass loss rates caused by a stellar wind instead of Roche lobe overflow)}{where the mass loss is due to stellar winds instead of RLOF}, we find a steady increase of the maximum temperature as a result of core growth due to helium and carbon burning, until conditions are met for neon ignition. In that case, neon ignition happens at a radial mass coordinate of $0.95 \msun$ for a core of $\replaced{1.375}{1.38} \msun$ with ignition closer to the center for higher core masses. }

\deleted{The red lines in Fig.~\ref{fig:tcrhoc_mloss-late} show the evolution of the core of the primary star in a binary with initial masses of $13.8 \msun$ (left panel) and $15.0 \msun$ (right panel). Both models fail to ignite neon, even though their \added{CO} core masses after carbon burning are $1.38\deleted{9} \msun$ and $1.51\deleted{4} \msun$ respectively, clearly above the canonical neon ignition threshold \citep{nom84e, jhn+13, tyu13, sqk16}. \deleted{The conditions at the location of maximum temperature reach a maximum of $\log T/\mathrm{K} = 8.97$ and $\log T/\mathrm{K} = 9.09$, and in both cases the temperature decreases again, even though the $15.0 \msun$ model might have developed a small amount of neon burning as it crossed the line where the neon energy generation rate exceeds the neutrino loss rate.} Over-plotted in blue are lines that show the evolution of a single star with an initial mass of $12.9 \msun$ (left panel) and $14.42 \msun$ (right panel). These masses have been chosen in such a way that the \added{mass of the} CO core right before central carbon burning is \replaced{similar}{identical} to that of the binary models ($1.28\deleted{4} \msun$ and $\replaced{1.395}{1.40} \msun$ respectively). Both single stars develop off-center neon burning \deleted{when the conditions at the temperature maximum exceed the line where $\epsilon_\mathrm{Ne} = \epsilon_\mathrm{\nu}$}. Even though the conditions prior to carbon burning were similar between the single and binary stars, and the final $M_\mathrm{CO}$ are \replaced{quite close}{identical} too, \deleted{(left panel: $1.389 \msun$ (single) vs. $1.391 \msun$ (binary); right panel: $1.514 \msun$ (single) vs. $1.515 \msun$ (binary))} the subsequent evolution of the core is \deleted{significantly} impacted by the high mass \replaced{loss}{transfer} rates due to RLOF observed in our models. \replaced{It causes, in the binary models, the initial increase of the maximum temperature as a result of core contraction to stop and reverse, leading to a cooling core, not only in the center, but also at the location of the maximum temperature, preventing conditions to develop under which neon can ignite.}{As a result, in binary models the initial increase in central temperature due to core compression reverses and is followed by the cooling of the entire core. This necessitates higher initial masses to allow for the development of neon burning (see Sec.~\ref{sec:neon-ignition}).}}

\begin{figure*}[tbp]
\begin{center}
\includegraphics[scale=0.58]{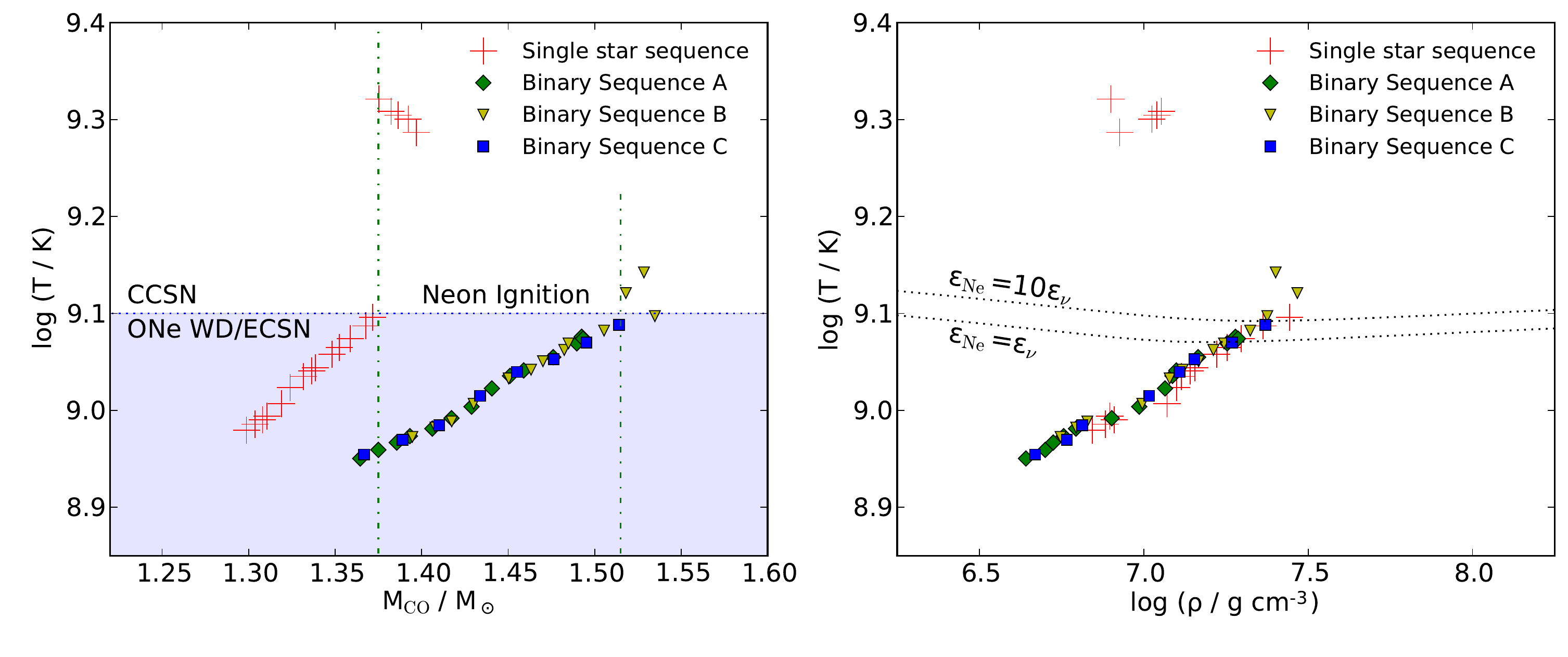}
\deleted{\caption{Maximum temperature attained in the core as a function of $M_\mathrm{CO}$, for single stars (red crosses) and primary stars in a binary system (filled symbols). The dotted line identifies the approximate position where the neon energy generation rate is equal to the neutrino loss rate. Stars with a maximum temperature above this line experience neon burning. Single stars have initial masses from $ 12.0 - 13.0 \msun$ (spaced by $0.05\msun$). Binary Sequence A consists of stars with the following initial parameters: $M_1 = 13.8 - 15.2\msun$ (spaced by $0.1 \msun$), $q=M_2/M_1 = 0.8$, and $P_\mathrm{init}=15$ days. Binary Sequence B consists of stars with the following initial parameters: $M_1 = 13.5 - 15.5\msun$ (spaced by $0.1 \msun$), $q=M_2/M_1 = 0.9$, and $P_\mathrm{init}=20$ days. Binary Sequence C consists of stars with the following initial parameters: $M_1 = 13.6 - 15.2\msun$ (spaced by $0.1 \msun$), $q=M_2/M_1 = 0.8$, and $P_\mathrm{init}=9$ days. All sequences have a mass transfer efficiency of $\beta = 0.5$. The correlation between the maximum attained temperature and $M_\mathrm{CO}$ is very different for single stars and binary components, but the period and mass ratio do not seem to affect the correlation. }}
\added{\caption{Maximum temperature attained in the core as a function of $M_\mathrm{CO}$ (left panel) and as a function of the density at the mass coordinate of $T_\mathrm{max}$ at the time of $T_\mathrm{max}$ (right panel), for single stars (red crosses) and primary stars in a binary system (filled symbols). The dotted line in the left panel approximates the critical temperature for neon ignition, while the dotted lines in the right panel give the locations where the energy generation of neon burning is equal to, and 10 times exceeds the energy losses due to thermal neutrinos.}}
\label{fig:mcore_maxT}
\end{center}
\end{figure*}

This difference in minimum $M_\mathrm{CO}$ needed for neon ignition between single stars and binary stars is also seen in Fig.~\ref{fig:mcore_maxT} \added{(left panel)} where we plot the final $M_\mathrm{CO}$ against the maximum temperature that a core attained during its evolution. The horizontal dotted line shows the approximate temperature that is necessary for neon to ignite\added{, and forms the boundary between CCSN and ECSN}. For simplicity we take a value of $\log T/\mathrm{K} = 9.1$, although this might vary slightly depending on the density as the right panel shows. The red crosses represent single star models with initial masses from $ 12.0 - 13.0 \msun$ (spaced by $0.05\msun$). These cross the $\epsilon_\mathrm{Ne} = \epsilon_{\nu}$ line at $M_\mathrm{CO} = 1.37\msun$, in good agreement with \citet{nom84e,jhn+13, tyu13, sqk16}. The filled symbols\deleted{, however,} represent several binary model runs. Binary Sequence A consists of stars with initial parameters: $M_1 = 13.8 - 15.2\msun$ (spaced by $0.1 \msun$), $q=M_2/M_1 = 0.8$, and $P_\mathrm{init}=15$ days. Binary Sequence B consists of stars with initial parameters: $M_1 = 13.5 - 15.5\msun$ (spaced by $0.1 \msun$), $q=M_2/M_1 = 0.9$, and $P_\mathrm{init}=20$ days. Binary Sequence C consists of stars with initial parameters: $M_1 = 13.6 - 15.2\msun$ (spaced by $0.1 \msun$), $q=M_2/M_1 = 0.8$, and $P_\mathrm{init}=9$ days. All sequences have a mass transfer efficiency of $\beta = 0.5$. \added{They all converge on one line, and as can be seen in the left panel of Fig.~\ref{fig:mcore_maxT}, the slope of this line is lower, and the maximum mass for neon ignition is shifted to considerably higher $M_\mathrm{CO}$ compared to the single star models.}\deleted{and are shifted to considerably higher $M_\mathrm{CO}$ and}\added{This line} only crosses the $\epsilon_\mathrm{Ne} = \epsilon_{\nu}$ line at $M_\mathrm{CO} = 1.52\msun$.\deleted{Binary models that do show neon ignition, show a much more moderate temperature increase than their single star counterparts ($0.05$ dex for binary stars versus an increase of $0.2$ dex for single stars, cf. red crosses around $\log\ T/\mathrm{K} = 9.3$ ).} \replaced{It appears}{The conclusion is} that stars in binaries, experiencing high RLOF mass transfer rates, need $M_\mathrm{CO} > 1.52 \msun$ to be able to \replaced{maintain a sustained temperature increase}{develop conditions} that will lead to neon ignition. \added{In the right panel of Fig.~\ref{fig:mcore_maxT} we plot the maximum temperature attained by the binary sequences and the single star sequence on a $T-\rho$ diagram. The fact that all sequences converge onto one line is compelling evidence that the CO cores in binaries behave physically the same, but undergo much stronger cooling than their single star counterparts, resulting in higher $M_\mathrm{CO}$ needed to ignite neon. Alternatively one could argue that these cores have, as a result of RLOF contracted to such high densities already, that further contraction is not able to raise the temperature in their cores to levels that could start neon ignition.} To this effect, the neon ignition boundary shown in Fig.~\ref{fig:ECSN_panel}  and Fig.~\ref{fig:ECSN_range} is positioned at a $M_\mathrm{CO} = 1.52\msun$, which we take as the boundary between possible ECSN progenitors (lower masses) and CCSN progenitors (higher masses). \added{See Appendix~\ref{sec:other_mixing} for evidence that this shift in critical $M_\mathrm{CO}$ for neon ignition is independent of the adopted treatment of convective boundaries or the inclusion of overshooting.}

\added{\section{The Evolution of the Cooling Core Towards ECSN}}\label{sect:new_ECSN_pathway}
\added{As argued by \citet{tyu13}, core contraction leading up to an ECSN goes through four distinct phases: First, neutrino cooling; second, core mass growth; third, electron captures on $^{24}$Mg and $^{20}$Ne and finally O+Ne deflagration. As our models show much stronger cooling than single star models, the question is how this will affect the subsequent evolution toward electron captures. Based on our models we see three possible scenarios, which all could be realized in various systems, depending on the initial conditions. 

The first possibility is based on scenario proposed by \citet{pac71} and further developed by \citet{sqb15, bbs+16}. In this scenario, the ONe core continues to increases in density and electrons become increasingly degenerate. As a result, the evolution of the core is more and more dominated by the balance between compressional heating and neutrino cooling. Compressional heating is a result of the gravitational contraction and core growth due to helium and carbon shell burning, while neutrino cooling is primarily due to thermal neutrino emission. As the density increases and the core temperature decreases, cooling due to plasma-neutrinos becomes less efficient, and eventually compressional heating slows down the cooling of the core. At some point in time, the neutrino cooling timescale, $\tau_\mathrm{\nu} = c_\mathrm{p}T/\epsilon_\mathrm{\nu}$, equals the compression timescale, $\tau_\mathrm{comp} = (\mathrm{d} \ln \rho / \mathrm{d}t)^{-1}$, and the evolution of the core will proceed along a trajectory which is primarily dependent on the core growth rate \citep{pac71, bbs+16}. When high enough densities are reached, URCA processes will accelerate the contraction of the core toward densities where electron captures on $^{20}$Ne and $^{24}$Mg will induce core collapse.

The second possible evolutionary scenario is prompted by some indications in our models that the primary enters a final mass transfer episode. Many of our models developed significant envelope instabilities during the final stages of their evolution (case ABC or BBC), which drove mass transfer rates up considerably. It is possible that this final episode of mass transfer removes the entire envelope, after which the remnant shrinks and mass transfer ceases, leaving a massive ONe core with possibly a thin CO shell around it. As this core is not able to grow anymore it will cool down and leave a massive, super-Chandrasekhar ONe WD. Evidence for super-Chandrasekhar white dwarfs comes from several peculiar Type Ia supernovae, e.g. SN 2006gz, SN 2007if, SN 2009dc \citep{hgp+07, saa+10, tbc+11} and theoretical considerations into their existence (e.g., \citealt{dm13, sm15}, for an arguments against their existence, see \citealt{mhp15}). To account for these super-Chandrasekhar cores, however, a strong magnetic field is required, and as we did not take magnetic fields into account in our models, we are not able to verify the existence of magnetic fields in these massive ONe cores.

The third possible scenario also takes its cues from the final mass transfer episode, but suggests that the intensity of the mass transfer will lead to the secondary filling its Roche lobe, leading to common envelope evolution. If this scenario develops, the binary components will enter a spiral-in phase with the orbital period significantly decreasing. The outcome of this scenario cannot be predicted at this time.

\begin{figure*}[tbp]
\begin{center}
\includegraphics[scale=0.58]{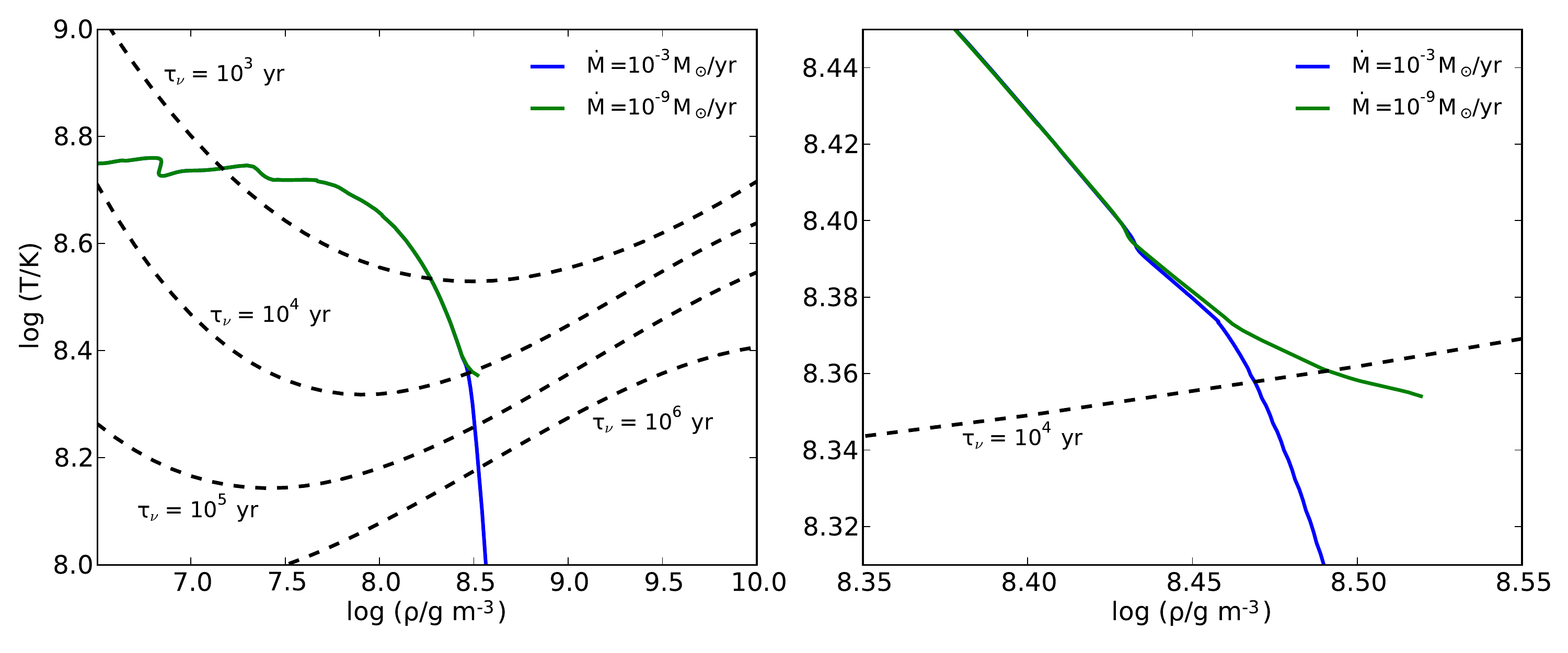}
\caption{Evolution of the ONe core in the $\rho_\mathrm{c}-T_\mathrm{c}$-plane. The right panel is a zoom in of the relevant part of the left panel. The solid lines represents models with a constant mass loss rate of $\dot{M} = 10^{-3} \msun$/yr (blue line) and $\dot{M} = 10^{-9} \msun$/yr (green line). The dashed lines indicate constant neutrino cooling timescale, with values given in the figure. The model with the low mass loss rate (green line) shows a hint of converging onto a common $\rho - T$ line where neutrino cooling and heating due to gravitational contraction are in balance.}
\label{fig:tcrhoc_attractor}
\end{center}
\end{figure*}

To investigate in more detail the final phase in the lives of our stars, and to establish evidence for any of these three scenarios, we continued several of our models beyond the endpoint of our regular grid (see Sect.~\ref{sect:grid} for details). All models that we evolved further, developed significant envelope instabilities during the final stages of their evolution (case ABC or BBC), leading to high mass transfer rates. While the reliability of these high rates can be questioned, the behavior itself is characterized by periodic variations in luminosity, stellar radius and hence mass loss rate, and has many similarities with the instabilities observed in models of red super giants \citep{hjl+97}. This final mass loss episode is very short and removed all or almost all of the remaining helium envelope, leaving a bare CO shell around a massive ONe core. The period of the oscillations we found is in the order of 4-8 years, which is in good agreement with the findings of \citealt{yc10, mesa2}. However, the instabilities can have either a physical or numerical origin \citep{lgd+12}, and assessing their validity necessitates a comprehensive resolution study, requiring a substantial number of models each consisting of a large number of time steps. As this kind of resolution study is outside the scope of this paper, we recomputed the late stages of several of our primaries by taking them outside of the binary and including an artificial viscosity term which allows for damping of the pulsations \citep{mesa3}. We also applied several different fixed mass loss rates to the models, to simulate a variety of possible mass transfer rates that could be attained in this final RLOF episode. All of these models, however, also ran also into instabilities, albeit at a higher density than the instabilities that occurred when they were still in the binary. Although we got an indication of their subsequent evolution, we were not able to follow the core evolution all the way to high enough densities for the onset of the URCA process or electron captures to be found.

Fig.~\ref{fig:tcrhoc_attractor} shows the evolution of two identical cores in a $\rho_\mathrm{c}-T_\mathrm{c}$ diagram, one with a fixed mass loss rate of $\dot{M} = 10^{-3} \msun$/year and the other with $\dot{M} = 10^{-9} \msun$/year. The dashed lines are lines of constant neutrino cooling timescales, calculated using the \texttt{pyMESA} package \citep{far17}. As can be seen, the model with a high mass transfer rate of $\dot{M} = 10^{-3} \msun$/year continues its cooling trend, as it loses its entire envelope and even eroding the core, preventing any further core growth and possible ensuing contraction. This led to a massive ONe WD with a mass of 1.42 $\msun$. The model with a low mass transfer rate of $\dot{M} = 10^{-9} \msun$/year slows down its cooling and gives an indication that it might converge onto a common $\rho - T$ line (\citealt{pac71}, \citealt{nom84e, nom87c} and also observed in Fig. 4 in \citealt{tyu13}), somewhere between the $\tau_\mathrm{\nu} = 10^4$ yr and $\tau_\mathrm{\nu} = 10^5$ yr line. The expectation is that this model will continue to contract, reaching higher and higher densities, aided by the URCA process, eventually leading to conditions that are conducive for electron captures to occur. This path will depend on the core growth rate, as argued by \citet{bbs+16}, and only detailed stellar models, which are able to avoid the discussed envelope instabilities, will provide us with clarity regarding this scenario.

As this scenario is the only course of events that will lead to an ECSN, we will adopt this to determine the initial conditions of close binaries that will produce ECSN. As the Chandrasekhar mass forms the mass limit for stable white dwarfs, we could take $M_\mathrm{Ch} = 1.4 \msun$ as the lowest possible mass for ECSN. However, our models do not give any indications that cores between $1.38 \msun$ and $1.4 \msun$ could not undergo this scenario, so in the most optimistic case, the minimum $M_\mathrm{CO}$ for ECSN is $1.38 \msun$, identical to the canonical limit for ECSN in single stars. The upper mass limit for ECSN is given by $M_\mathrm{CO} = 1.52 \msun$, as argued in Sec.~\ref{sec:late_mass_loss}.

\section{Progenitor Structure}
The progenitor structure provides significant information regarding the explosion dynamics and the weak and nuclear reactions which can occur in the event of a supernova explosion. The top four panels in Fig.~\ref{fig:progenitor_structure} show the chemical structure of our Case A model (Sec.~\ref{sec:Case_A}) and our early and intermediate Case B models (Sec.~\ref{sec:Case_B}). In addition we also provide the chemical structure of a model with a super-Chandrasekhar CO core ($M_\mathrm{CO} = 1.51 \msun$), which evolved from the initial parameters $M_1 = 15.0\msun$, $M_2 = 12.0 \msun$,  $P_\mathrm{init} = 9$ days and $\beta = 0.5$. All models show a ONe core with a small amount of Mg. The $^{12}$C layer is very thin and forms the transition from the dense core to the low density envelope. The density structure is shown in the bottom two panels and shows a steep density gradient at the transition from the core to the envelope. This steep gradient resembles the steep gradient that is characteristic of SAGB models (c.f. \citealt{jhn+13}, Fig. 7). Temperature profiles are also plotted and show an off-center temperature maximum at a radial mass coordinate close to the carbon burning shell (indicated with a green diamond), indicating that the entire core is cooling.

If these cores are able to explode, the steep density gradients separating the core and the envelope will give rise to a relatively weak and fast explosion,  producing a small natal kick, compared to the large natal kick (in the order of 400-500 km s$^{-1}$) experienced after stronger and slower CCSN  \citep{kjh06, jhh+12b, jan17}. This dichotomy between small and large natal kick could possibly explain the bimodal distribution in spin period and orbital eccentricity of X-ray binaries \citep{plp+04, kcp11}. It is also relevant in the context of double neutron star mergers, where the system remains bound even after experiencing two supernova explosions. This requires at least a small kick in the second supernova, but possibly also significantly smaller kicks in the first supernova, which appears at least possible given the density structure of our models \citep{bp16, tkf+17}.

\begin{figure*}[tbp]
\begin{center}
\includegraphics[scale=0.58]{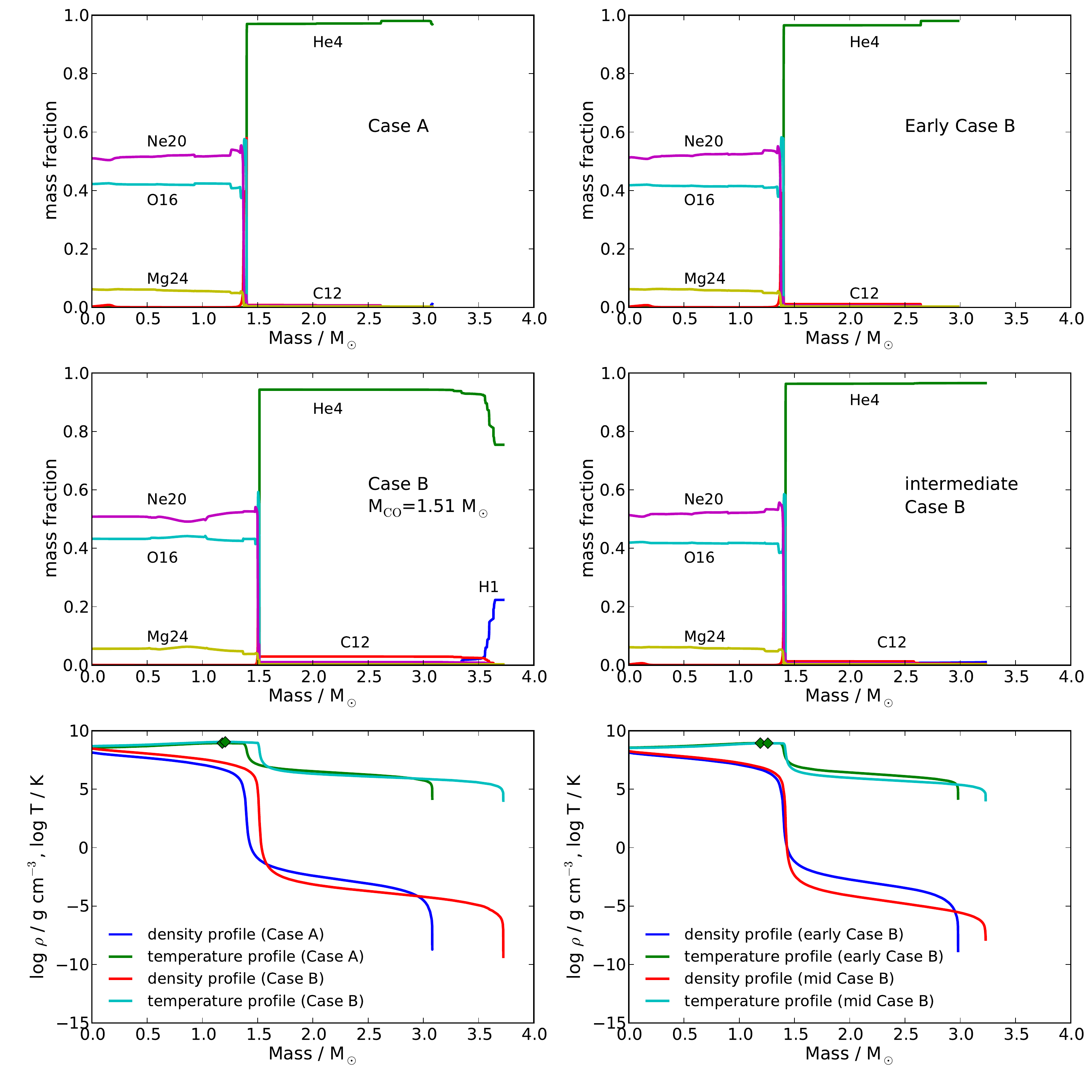}
\caption{Progenitor structure of four different models. The top rows show the chemical structure as a function of the mass coordinate. The bottom row shows the temperature and density structure of the progenitor model as a function of mass coordinate. All models posses an ONe core, with a fraction of Mg. The structure resembles that of a single SAGB star with a sharp density gradient separating the core from the envelope, c.f. \citealt{jhn+13}, Fig. 7. The temperature profiles show an off-center maximum, marked with a green diamond.}
\label{fig:progenitor_structure}
\end{center}
\end{figure*}
}

\deleted{\subsection{The Final Evolution of Select Systems}\label{sec:final_evolution}
We studied in more detail the final evolution of a particular system ($M_1 = 13.7\msun$, $M_2=10.96\msun$, initial mass = $9$ days, $\beta$ =  0.5) as far as our simulations would allow. This system is representative of a early Case B system with mass loss starting $18,000$ years after hydrogen core burning finished. The system evolves similarly to the early Case B system that was described in Sec.~\ref{sec:Case_B} with a total mass of the primary at the end of central carbon burning of $2.96\msun$. The system undergoes two periods of mass transfer. The first phase lasts for $27,200$ years and after an initial peak with a mass loss rate of $3.35 \times 10^{-3} \msun$\,yr$^{-1}$ the mass loss slows down to more moderate rates ($4.0 \times 10^{-5} \msun$\,yr$^{-1}$). During this Case B mass loss episode the star loses $10.3\msun$. A second distinct phase starts as the envelope expands after the termination of core carbon burning. This phase lasts for $20,000$ years and proceeds at a more moderate rate of $\sim 10^{-5}\msun$\,yr$^{-1}$. During this phase the star loses another $0.25\msun$. The final mass loss episode lasts for about a $70$ years and removes almost all of the remaining helium envelope (see Fig.~\ref{fig:final_evolution}). This mass loss episode is characterized by periodic variations in luminosity, stellar radius and therefore mass loss rate, with rates briefly increasing to $\sim 10^{2}\msun$\,yr$^{-1}$. While the reliability of these rates can be questioned, the phase itself has many similarities with the instabilities observed in models of red super giants \citep{hjl+97}. In both cases, the period of the oscillations is in the order of 4-8 years, which is in good agreement with the findings of \citealt{yc10, mesa2}. 

\begin{figure*}[tbp]
\begin{center}
\includegraphics[scale=0.56]{f10_old.pdf}
\caption{Final (and fatal) mass loss episode of a system with initial parameters of $M_1 = 13.7\msun$, $M_2=10.96\msun$, initial period = $9$ days, and $\beta =  0.5$. Top Left: Evolution of the mass loss rate as a function of time.  Severe instabilities in the envelop of the star cause sudden increases in the mass loss rate, with a period that approaches $4$ years. Top Right: Evolution of the total mass of the donor ($M_1$) and the gainer ($M_2$, right alternate axis) and the carbon-oxygen ($M_\mathrm{CO}$) and oxygen-neon ($M_\mathrm{ONe}$) core of the donor. As the donor loses its mass, the gainer receives half of it ($\beta = 0.5$). Bottom Left: Evolution of the luminosity as a function of time. Bottom Right: Evolution of the stellar radius (red), Roche radius (bottom blue) and orbital separation (top blue, right alternate axis). During the final mass loss episode the orbital separation grows from $300 \rsun$ to $1030 \rsun$. Shortly after this mass loss episode the model terminates due to instabilities.}
\label{fig:final_evolution}
\end{center}
\end{figure*}

Mass and angular momentum loss drives the widening of the orbit from $300 \rsun$ to $1030 \rsun$.  During these instabilities the star loses almost its entire envelope, leaving eventually a helium layer of $0.0578 \msun$, and a carbon rich layer of $0.0364 \msun$. Because of instabilities in the envelope we were not able to continue the calculation beyond this point, but assuming that the mass loss rate stays at a level of $\sim 1 \times 10^{-5} \msun$\,yr$^{-1}$ the star will lose its remaining helium in less than $6000$ year. This will terminate any residual helium burning, and most likely also carbon burning, and the star will continue to cool down through core contraction and forma a super-Chandrasekhar mass ONe white dwarf. Its final fate is not totally clear as sustained contraction will continue to increase the central density. Since neutrino cooling becomes less important for temperatures below $1 \times 10^8$ K \citep{ihh+96}, the cooling of the core should slow down and all ONe cores in this situation should converge onto a common evolutionary track as suggested by \citet{pac71} (see also Fig. 4 in \citealt{tyu13} which gives evidence of this common evolutionary track). This might eventually lead to electron captures on $^{20}$Ne and $^{24}$Mg and result in an ECSN, but we are unable to determine the exact evolution of the stars, nor its final fate, because of instabilities in the envelope.}

\section{The Mass Range for ECSN}\label{sec:ECSN_mass_range}
\begin{figure}[tbp]
\begin{center}
\includegraphics[scale=0.57]{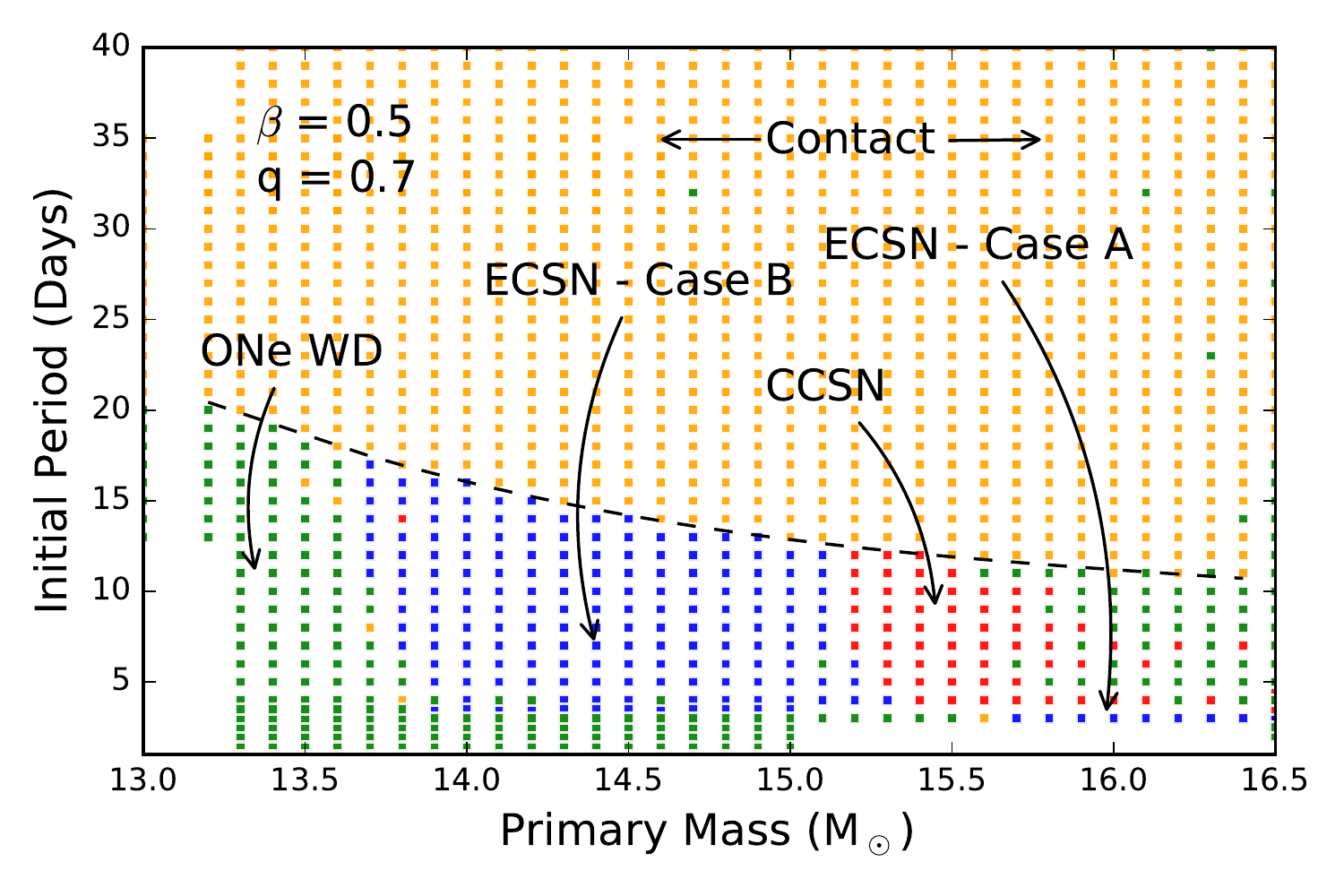}
\caption{Expected final fate of computed models in our grid for $\beta = 0.5$ and $q = 0.8$ as a function of initial primary mass and initial period. Five distinct outcomes are visible and labeled. ECSN are indicated by blue dots.\deleted{Five distinct regions are visible. Primaries that evolve into a massive ONe WD are indicated by green squares. Contact systems are indicated with yellow squares and primaries that develop neon burning and will evolve toward an CCSN are indicated with red squares. ECSN progenitors are indicated with blue squares. The Case B ECSN progenitor sequence is visible between initial masses of $13.7\msun$ and $15.2\msun$, while primaries that undergo Case A Roche lobe overflow and develop a CO core in the ECSN mass range are visible at an initial period of $3$ days with initial masses $>15.7\msun$.}}
\label{fig:ECSN_panel_one}
\end{center}
\end{figure}

\begin{figure*}[htbp]
\begin{center}
\includegraphics[scale=0.53]{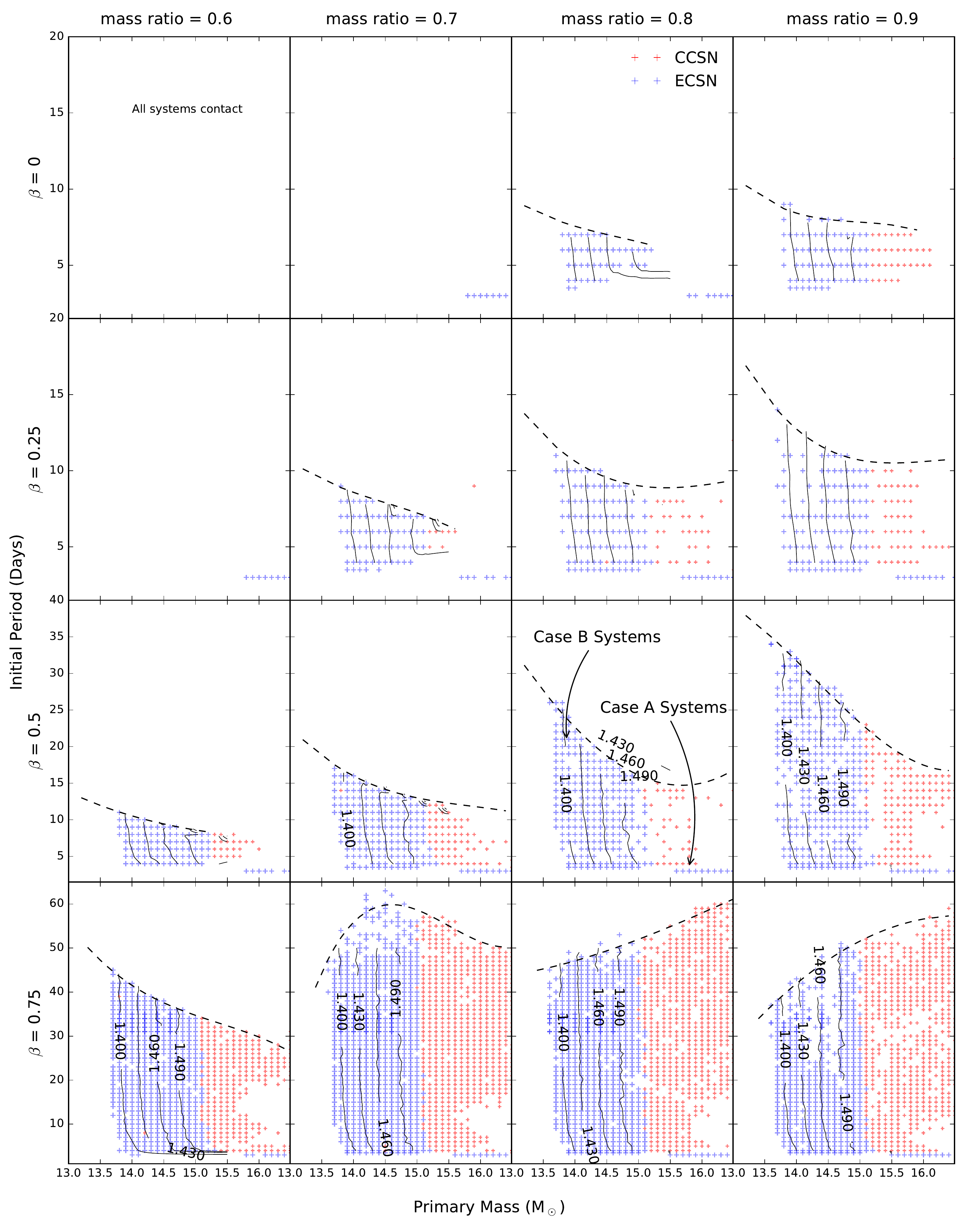}
\caption{\replaced{Models that lead to a supernova}{ECSN (blue) and CCSN models (red)} as a function of initial primary mass and initial period, for $\beta = 0, 0.25, 0.5, 0.75$ and mass ratios $q = 0.6, 0.7, 0.8, 0.9$.\deleted{Electron capture supernovae are indicated by blue crosses, core collapse supernovae are indicated by red dots. Systems above the dashed line are contact systems, but for clarity sake we did not plot them (i.e., the dashed line indicates the lowest initial period for which systems develop contact). We also did not plot systems that develop a core below $1.37\msun$ -- they reside to the left of, and below the ECSN progenitors.}The initial period for the two top rows is limited to $20$ days, and the third row to $40$ days as systems with larger periods evolve all to contact binaries. Case A and Case B systems are indicated in the panel for ($q$,$\beta$) = (0.8, 0.5) and are representative for all other panels. Masses of the pre-ECSN CO cores are indicated with contour lines, marking the location of $1.40, 1.43, 1.46$, and $1.49 \msun$. The approximate boundary between contact and contact-free systems is indicated by a dashed line.\deleted{Case A systems can seen to be shifted to higher initial masses. CO core masses are indicated with contour for masses of $1.40, 1.43, 1.46$, and $1.49 \msun$. For $\beta = 0$ (fully conservative evolution, top row) ECSN occur only for systems with mass ratios close to $1$ and for low periods ($P_\mathrm{init} < 10$d). For higher values of $\beta$ (increasingly more mass is lost from the system) the range for which ECSN occur shifts to slighly lower primary masses, but includes a wider range of mass ratios ($0.6 < q < 1$) and initial periods, up to $P_\mathrm{init} = 60$d.}}
\label{fig:ECSN_panel}
\end{center}
\end{figure*}

Based on our results described above, we are now in the position to discuss the parameter space where conditions for ECSN are favorable.
A total of approximately 45,000 binary sequences were \replaced{run}{calculated} to investigate the ECSN channel in binary stars. A small fraction of the models (approximately $2000$) \added{the majority with initial conditions that bring them close to evolving into a contact system,} suffered numerical instabilities during \replaced{the early stages of the}{their} evolution and were terminated because of that. These models were ignored in the final analysis. Most models, however, capture the evolution of the stellar models through carbon burning and the formation of an ONe core, unless the system developed contact. The final fate of our models can therefore be characterized by \replaced{five}{six} different outcomes, which are summarized below and shown in Fig.~\ref{fig:ECSN_panel_one} for one particular combination of parameters, i.e. $\beta = 0.5$ and $q=0.7$. The full grid is shown in Fig.~\ref{fig:ECSN_panel} and will be discussed in more below.
\begin{itemize}
\added{\item (Short-P-contact) Models with a short initial period ($P_\mathrm{init} \lsim 2.5$ days at $q = 0.9$ and decreasing to $P_\mathrm{init} \lsim 1.5$ days at $q = 0.7$) will develop a contact system as the primary fills its Roche lobe during the main sequence and subsequent mass transfer shrinks the orbit so much that also the secondary fills its Roche lobe and contact ensues. These models are not shown in Fig.~\ref{fig:ECSN_panel_one}, but the results agree with \citet{wlb01}.}
\item (\replaced{High}{Long}-P-contact) Binary systems with a \replaced{high}{long} initial period will develop contact as well, due to the high mass transfer rates which prevent the secondary to adjust sufficiently fast enough to avoid filling its Roche lobe (indicated with yellow symbols). The late Case B model described in Sec.~\ref{sec:Case_B} is part of this class.
\item (ONe WD) Models with a low initial primary mass ($\lsim 13.5\msun$) will develop a massive ONe WD (indicated with green symbols).
\item (CCSN) Models with a high initial primary mass ($\gsim 15.2\msun$) will develop neon burning and will evolve toward a CCSN (indicated with red symbols).
\item (Case B ECSN) Models with a period between $\sim \replaced{3}{4}$ days and an upper bound that depends on the values of $\beta$ and $q$ and initial primary masses of $\sim 13.5 - 15.2 \msun$ develop Case B mass transfer and a CO core that falls with a mass between $1.37\msun$ and $1.52\msun$ (indicated with blue symbols, ECSN - Case B). 
\item (Case A ECSN) Models with an initial period that is \replaced{around}{between} $\sim \replaced{3}{2.5}$ \added{and $\sim 3.5$} days \added{(depending on the value of $q$)} will develop late Case A mass transfer. \added{As the development of the helium core is significantly hampered by this mass transfer, the} primary initial mass range shifts \added{to higher initial masses} with respect to Case B systems\deleted{because of core mass reduction due to early mass loss}(indicated with blue symbols, ESCN - Case A). 
\end{itemize}

In Fig.~\ref{fig:ECSN_panel_one} we show these final evolutionary outcomes for a particular combination of parameters, i.e. $\beta = 0.5$ and $q=0.7$ (compare also with \citealt[][Fig. 12]{wlb01}). \added{As argued above, while in single stars CO cores with masses between $1.37 \msun$ and $1.42\msun$ will result in an ECSN, in binary stars these ranges could possibly be as wide as $1.37 - 1.52\msun$.} ECSN progenitors have a minimum mass of $13.7\msun$ which increases to $13.9\msun$ for early Case B systems (\replaced{analogical}{analogical} to the much \replaced{bigger}{larger} shift in initial mass for Case A systems). The upper mass \replaced{bound}{limit} for ECSN progenitors is found at $15.1\msun$ with a slight shift to higher initial masses ($15.3\msun$) for early Case B systems. This boundary forms the transition to progenitors that develop neon burning. Case A ECSN progenitors can be found at $P_\mathrm{init} = 3$ days, between $M_1 = 15.7\msun$ and $M_1=16.4\msun$. For this particular combination of $q$ and $\beta$, the minimum period for systems to develop contact decreases from about $20$ days at $M_1 = 13.5\msun$ to $\sim 10$ days at $M_1=16.5\msun$. This limits the number of ECSN candidates at higher initial masses, but also the progenitors that develop neon burning.  

\deleted{Figure~\ref{fig:ECSN_panel} shows ECSN progenitors of our entire parameter space as a function of initial primary mass and initial period, with four different values of the mass ratio $q$ plotted horizontally ($q = 0.6, 0.7, 0.8, 0.9$), and four different values of $\beta$ plotted vertically. CO core masses between $1.37$ and $1.52$ are indicated with contour lines.}

\added{Plotted in Fig.~\ref{fig:ECSN_panel} are all ECSN and CCSN progenitors (respectively red crosses and blue crosses) in our entire parameter space as a function of initial primary mass and initial period, with four different values of the mass ratio $q$ plotted horizontally ($q = 0.6, 0.7, 0.8, 0.9$), and four different values of $\beta$  plotted vertically \added{($\beta = 0, 0.25, 0.5, 0.75$)}. The approximate boundary between contact and contact-free systems is indicated by a dashed line. This line is a third order polynomial fit to the maximum contact-free period, and hence should be taken as an approximation. $M_\mathrm{CO}$ between $1.37 \msun$ and $1.52 \msun$ are indicated with contour lines at $M_\mathrm{CO} = 1.4, 1.43, 1.46, 1.49 \msun$.} 

As can be seen, most combinations of $q$ and $\beta$ behave qualitatively similar\added{, especially with respect to the position of the initial mass limits for ECSN}, although the boundaries between contact systems and non-contact systems \replaced{differs}{shift for various combinations of $q$ and $\beta$}. \added{A general trend (except for $\beta = 0.75$ with $q = [0.7, 0.8, 0.9]$) can be seen such that for lower initial primary masses contact-free systems can be found toward longer initial periods. This is the result of lower mass transfer rates due to RLOF at lower masses (c.f. Fig.~\ref{fig:CaseB-typical_combined}), which allows the secondary to adjust to the mass accretion, instead of overflowing its Roche lobe. For lower values of $q$ (i.e. a larger difference between the initial primary and secondary mass) we see the boundary between contact and non-contact systems shift to shorter period. As the mass difference is larger for low $q$, the size of the Roche lobe of the secondary is smaller, and once mass transfer starts, it is easier for the secondary to fill the Roche lobe and develop contact. This effect is stronger for more efficient mass transfer, as it is more difficult for the secondary to adjust its radius in response to the larger amounts of accreted mass. As can be seen in Fig.~\ref{fig:ECSN_panel}, upper left corner, systems with $q = 0.6$ and $\beta = 0, 0.25$  all develop contact. A clear trend is visible for less efficient mass transfer toward more contact-free systems at long periods. As more matter leaves the system, less matter is accreted onto the secondary, which does not fill its Roche lobe until longer periods.}\deleted{ (shifts to lower initial period for decreasing $q$ values, and to higher periods for increasing $\beta$ values). }

\added{For most combinations of $q$ and $\beta$, }the initial primary mass for ECSN is confined to a narrow range \replaced{from}{between} $13.6\msun$ \replaced{to}{and} $15.1\msun$ for Case B systems and between $15.5\msun$ and $17.6\msun$ for Case A. For fully conservative evolution ($\beta = 0$) the initial periods for systems Case B mass transfer that result in an ECSN are confined to $P_\mathrm{init} < 10$ days and mass ratios $q > 0.75$. Several Case A systems are \replaced{visible}{found} for initial primary masses between $15.8\msun$ and $16.4\msun$ at a period of $3.5$ days. \replaced{There are no }{All } Case B systems for mass ratios below $0.7$ \added{develop contact}, however, there are Case A systems with an initial period of $3$ days and initial primary masses between $15.8\msun$ and $16.3\msun$. \added{This is in good agreement with \citet{wlb01} who find that Case A systems avoid contact for lower $q$ compared with Case B systems.}

For non-conservative evolution the mass range for ECSN barely changes, however, the range in initial periods does increase. For $\beta = 0.25$ (25\% of the mass lost from the primary star is expelled from the system) the maximum period increases to about $\sim 15$ days for a mass ratio near unity and decreases to $10$ days for a mass ratio of $0.70$. \deleted{There are no Case B systems for mass ratios below $0.75$.}

For non-conservative evolution with $\beta = 0.5$ ($50$\% accreted, $50$\% expelled) the maximum initial period for Case B systems increases to $33$ days for $q=0.9$, to $25$ days for $q=0.8$ and to $17$ days for $q=0.7$. The systems discussed in Section~\ref{sec:rep_systems} are part of this data set. Case A mass transfer leading to ECSN occurs for all mass ratios, while Case B mass transfer is still limited to $0.7 < q < 1$. 

The situation for the non-conservative case with $\beta = 0.75$ ($75$\% expelled, $25$\% accreted) is a bit different as the maximum period for ECSN formation increases from $50$ days in the $q=0.9$ case to $57$ days in the $q=0.7$ case. Models with a value of $\beta = 0.75$ form the only instance for $q=0.6$ that shows ample evidence for Case B systems able to evolve to a ECSN. These Case B systems are found between $13.7\msun$ and $15.0\msun$ with an initial period between $4$ days and $45$ days.

\section{Discussion and Conclusions}\label{sec:discussion}
We have presented \added{models of} close binary \replaced{models}{systems} where the primary star could potentially be the progenitor of an ECSN. While the mass range for ECSN in single stars is fairly narrow \citep{phl+07}, the mass range for ECSN in binary stars is thought to be much wider as the effects of mass loss due to Roche lobe overflow are thought to mitigate the effects of the second dredge-up in single stars \citep{plp+04}. \added{In single stars, }this second dredge-up reduces the mass of the helium core below the Chandrasekhar mass (in the relevant initial mass range) and makes that the possibility for an ECSN depends on the outcome of the race between core-growth and envelope mass loss during the SAGB \citep{phl+07, tyu13}. If the core is able to contract to high enough densities so that electron captures on $^{24}$Mg and $^{20}$Ne can commence, heating as a result of these electron captures will cause O+Ne burning at the center, and O+Ne deflagration propagates outward. Core contraction is further accelerated by electron captures in the central nuclear statistical equilibrium region, resulting in a weak Type II supernova \citep{tyu13, kjh06}. However, if mass loss removes the envelope fast enough so that the core is not able to reach those critical densities, the star will evolve into a massive ONe white dwarf.

Our study, however, has focused on binary systems instead of single stars and attempts to answer two major questions concerning: First, are binary systems indeed capable of producing and ECSN, and, second, what is the expected region in the ($M_1$, $M_2$, $P_\mathrm{init}$) phase space that we can expect these ECSN to occur? We have investigated these questions by running approximately 45,000 binary models in the relevant phase space.

\subsection{The Possibility of ECSN from Close Binary Systems}
Based on single star models there is consensus in the literature that a star will experience an ECSN when $M_\mathrm{CO}$ is somewhere between $\sim 1.37$ and $\sim 1.42\msun$ \citep[see for example][who all adopt similar values]{wh15,tlp15,me16}. \citet{tlp15} \replaced{argue}{, in the context of ultra-stripped supernovae, use as a ``rule of thumb'' (based on single star models)} that the upper boundary \added{for ECSN} is given by stars that develop a post-carbon burning central temperature above their carbon burning central temperature as \replaced{they will experience}{these conditions will lead to} an iron core collapse. In this case the ignition of neon and oxygen burning will eventually convert the composition of the entire core into $^{28}$Si and $^{32}$S, bringing the chance for contraction due to electron captures to an end. Their lower boundary is given by the ONe WD threshold of $1.37 \msun$, as these cores are not able to contract to sufficiently high densities where electron captures can commence.

Based on the models presented above, however, we question whether these boundaries can be applied to stars in a binary system, as mass loss driven by Roche lobe overflow, especially after central carbon burning drives the star out of thermal equilibrium, leading to significant expansion and a much stronger cooling in the core than in single stars. While in single stars the effect of neutrino cooling is compensated by heating due to core growth, keeping the central temperature of the star roughly constant during the post-carbon burning contraction \citep{nom84e}, in binary stars the cooling is enhanced by mass loss.\deleted{In addition, instabilities that develop after carbon burning, drive very strong mass loss which removes the remaining helium envelope and inhibits the chance for any future core growth. As argued by \citet{tyu13}, contraction toward collapse goes through four distinct phases: First, neutrino cooling; second, core mass growth; third, electron captures on $^{24}$Mg and $^{20}$Ne and finally O+Ne deflagration. In the models presented above, the first stage, neutrino cooling is enhanced by expansion due to mass loss, while the second stage, core mass growth is prevented due to extreme mass loss removing the remaining envelope. Taken together, this paints a very gloom picture for the possibility of an ECSN from close binary systems. 

Evidence for super-Chandrasekhar white dwarfs comes from several peculiar Type Ia supernovae, e.g. SN 2006gz, SN 2007if, SN 2009dc \citep{hgp+07, saa+10, tbc+11} and theoretical considerations into their existence (e.g., \citealt{dm13, sm15}, for an arguments against their existence, see \citealt{mhp15}). To account for these super-Chandrasekhar cores, however, a strong magnetic field is required, and while we did not take magnetic fields into account in our models, we doubt that these CO cores have strong fields because of their slow rotation. We question, therefore, whether the formation of super-Chandrasekhar cores is tenable. On the other hand, it is entirely possible that, after these cores lose their envelope and mass loss ceases, contraction will accelerate again, maybe combined with some very limited CO or ONe core growth, which stops the temperature drop and leads the core onto a path toward higher densities and subsequent electron-captures again. If these cores, as a result of continued core contraction or core growth are indeed able to increase their central density, they will most likely eventually merge onto a common evolutionary track in the $\rho_\mathrm{c}-T_\mathrm{c}$ plane (\citealt{pac71}, \citealt{nom84e, nom87c} and also observed in Fig. 4 in \citealt{tyu13}). This path will be dependent on the core growth rate, as argued by \citet{bbs+16}, and only detailed stellar models are able to provide us with the order of magnitude of this core growth rate.} \added{However, if cooling due to thermal neutrino losses and heating due to gravitational contraction come into equilibrium as argued in Sec.~\ref{sect:new_ECSN_pathway}, the central conditions will eventually follow a common evolutionary track in the $\rho_\mathrm{c}-T_\mathrm{c}$ plane and the core will be able to evolve to conditions that are conducive for electron captures to commence.} Nevertheless, \deleted{evolution toward an ECSN is certainly possible, although }many ingredients for this contraction are different from the evolution of single stars, and a simple comparison of \added{single and binary} evolutionary tracks in the $\rho_\mathrm{c}-T_\mathrm{c}$ plane \added{to deduce their final fate} is not possible. This scenario would apply to all cores that do not ignite neon ($M_\mathrm{core} \lesssim 1.52 \msun$) down to the effective Chandrasekhar mass ($\sim 1.40 \msun$) or possibly even $\sim 1.37 \msun$ and deserves further investigation.

\deleted{\subsection{Indirect Scenario: Reversed Mass Transfer}\label{sec:indirect_scenario}}
\replaced{In another scenario, possibly an extension of the ``gloom'' scenario described above, }{In case} the primary star \added{loses its entire envelope and is not able to contract to densities high enough for electron captures to commence, there is still an alternative scenario that could lead to an ECSN.} \deleted{evolves into a ONe white dwarf with a mass above the Chandrasekhar mass. It has lost almost its entire envelope and the core is too cool to ignite neon, or not dense enough yet to allow electron captures to start. However,}Once the secondary leaves the main sequence, its expansion will give rise to reverse Roche lobe overflow \citep{dv03,zmi+17}. This will likely happen not immediately, as the binary separation has grown to $\sim 1000 \rsun$. Eventually, however, \added{only a tiny bit of} mass accretion onto the ONe white dwarf will heat up the cold white dwarf, allow core growth to resume, reheat the core and possibly give rise to conditions that are conducive for electron captures to accelerate the heating process. \added{This will ultimately, although delayed compared with the direct explosions described above,} result in the collapse of the core and the formation of a neutron star \citep{nom84e, dbo+06, sqk16}. \deleted{As we were not able to continue our models beyond the envelope instabilities mentioned in Sec.~\ref{sec:final_evolution} we were not able to evaluate the viability of this accretion-induced scenario.}

\subsection{Comparison with Previous Work}
The results presented in this paper differ significantly from earlier studies that evaluated the existence of ECSN in binary systems. The first paper to discuss such supernovae in binaries was \citet{plp+04} who concluded that, based on models by \citet{wlb01}\deleted{ and \cite{phl+07}}, initial primary masses between $8 \msun$ and $17 \msun$ would be expected to evolve into an ECSN. This estimate was based on the helium core criterion developed by \citet{nom84e} who showed that helium cores between $2\msun$ and $2.5\msun$ lead to conditions where electron captures will kick off the collapse of the core. 

Despite the differences, many of the fundamental ideas in \citet{plp+04} are confirmed by this paper. While their inference was based on only a handful of models that established the relationship between the initial mass and final helium mass of stars in close binary systems, our detailed grid confirms indeed that the spread in $M_\mathrm{CO}$ is fundamentally a result of the period, and that the difference in timing of Roche lobe overflow between Case A and Case B systems will result in Case A systems developing smaller cores, effectively shifting their initial mass-final core mass relationship toward higher initial masses.

Our research, however, provides several improvements on \citet{plp+04}. First of all, the helium core criterion, developed by \citet{nom84e} might work well for single stars, it, however, gives less accurate results for binary stars. The main reason for this is that \added{primaries in} close binary \replaced{stars}{systems} suffer from significant mass loss \added{due to RLOF}, either during the main sequence or in the Hertzsprung gap, which considerably affects the development and mass of the helium core\added{, either directly by erosion, or indirectly through the adjustment of the star to mass transfer}. \replaced{Our initial thoughts were that we could replace this helium core criterion with a CO core criterion, as the CO core is itself is not eroded through mass loss, and is only established after helium burning, when the most severe mass loss episode is over. }{Although the CO core is not directly affect (i.e. eroded) by mass transfer due to RLOF, there is still an appreciable difference between the evolution of CO cores of the same mass in single stars and in binary systems, as Fig.~\ref{fig:mcore_maxT} shows.}\deleted{However, as Figs.~\ref{fig:mcore_maxT} and \ref{fig:ECSN_range} show, even for CO cores there is a considerable difference between single stars and stars in a binary system.} Secondly, whereas \citet{plp+04} expected that mass loss due to Roche lobe overflow would prevent the second dredge-up from happening, we find that mass loss actually has a very similar effect in reducing the mass of the helium core. Indeed, the second dredge-up was avoided, but the mass of the hydrogen envelope and sometimes the underlying helium layer was significantly reduced. Evidence for this erosion of the helium layer was already present in \citet{wlb01}, however, it was not considered in \citet{plp+04}. Third, our models show that the role of mass loss in the final evolution toward electron captures is much larger than was previously thought. Cores undergoing significant mass loss compensate for this by expansion and enhanced cooling. This leads to massive ONe cores (up to $\sim 1.52 \msun$) that are able to avoid neon ignition.\deleted{On the one hand, this makes them excellent candidates for ECSN as their cores have not been processed by O+Ne burning, however, the rapid cooling together with the loss of the entire envelope in the final stages of their evolution, will prevent them from contracting to densities high enough for electron captures to commence.} If the ONe core\deleted{, however,} is able to contract to high enough densities to cause the conditions for electron captures to occur we expect a maximum mass range for Case B systems that is $\sim 2 \msun$ wide and a maximum mass range for Case A systems that is $\sim 3 \msun$ wide. When we combine both mass ranges, the maximum mass range for ECSN from binary systems runs from an initial primary mass of $\sim 13.5 \msun$ to $\sim 18 \msun$, a width of $\sim 4.5 \msun$, which is significantly narrower than the prediction of \citet{plp+04} but much wider than the initial mass range for single stars (roughly $0.25 \msun$, \citealt{phl+07}). The use of a different convection criterion (e.g. Schwarzschild instead of Ledoux, more efficient semi-convection, or additional convective boundary mixing), will translate the mass range to lower values (possibly somewhere around $\sim 10 - 15 \msun$), without affecting the primary conclusions of this paper (c.f., Appendix~\ref{sec:other_mixing}, \citealt{dsl+10}). This will be \replaced{the subject of}{further explored in} a future publication. Even so, the improvements of the models that we have presented there provide a picture that makes the possibility for ECSN from close binary systems much less likely than originally thought. 

Our models are in broad agreement with the results of \citet{tlm+13,tlp15}. While their research is focused on helium cores orbiting a compact object, the general evolutionary picture shows lots of similarities (compare Fig.~18 in \citealt{tlp15} with our Fig.~\ref{fig:ECSN_panel}). As it is not clear which role mass loss plays in their models, especially during and after carbon burning, we don't know whether our results are applicable to their situation. Regardless, more research is needed to accurately describe the evolution of massive ONe cores at high densities, whether they converge onto a common evolutionary track in the $\rho_\mathrm{c}-T_\mathrm{c}$ plane and are able to contract to sufficiently high densities for electron captures to destabilize the core, or that they continue to cool down and avoid electron captures all together. Much of this will determine the exact mass range of ECSN in binary systems as neither the mass of the CO core, nor the track in the $\rho_\mathrm{c}-T_\mathrm{c}$ plane, are, according to our models,  sufficient to uniquely determine the final fate of these stars.

\subsection{The Expected Initial Mass Range for ECSN in Close Binary Systems}
\begin{figure*}[htbp]
\begin{center}
\includegraphics[scale=0.58]{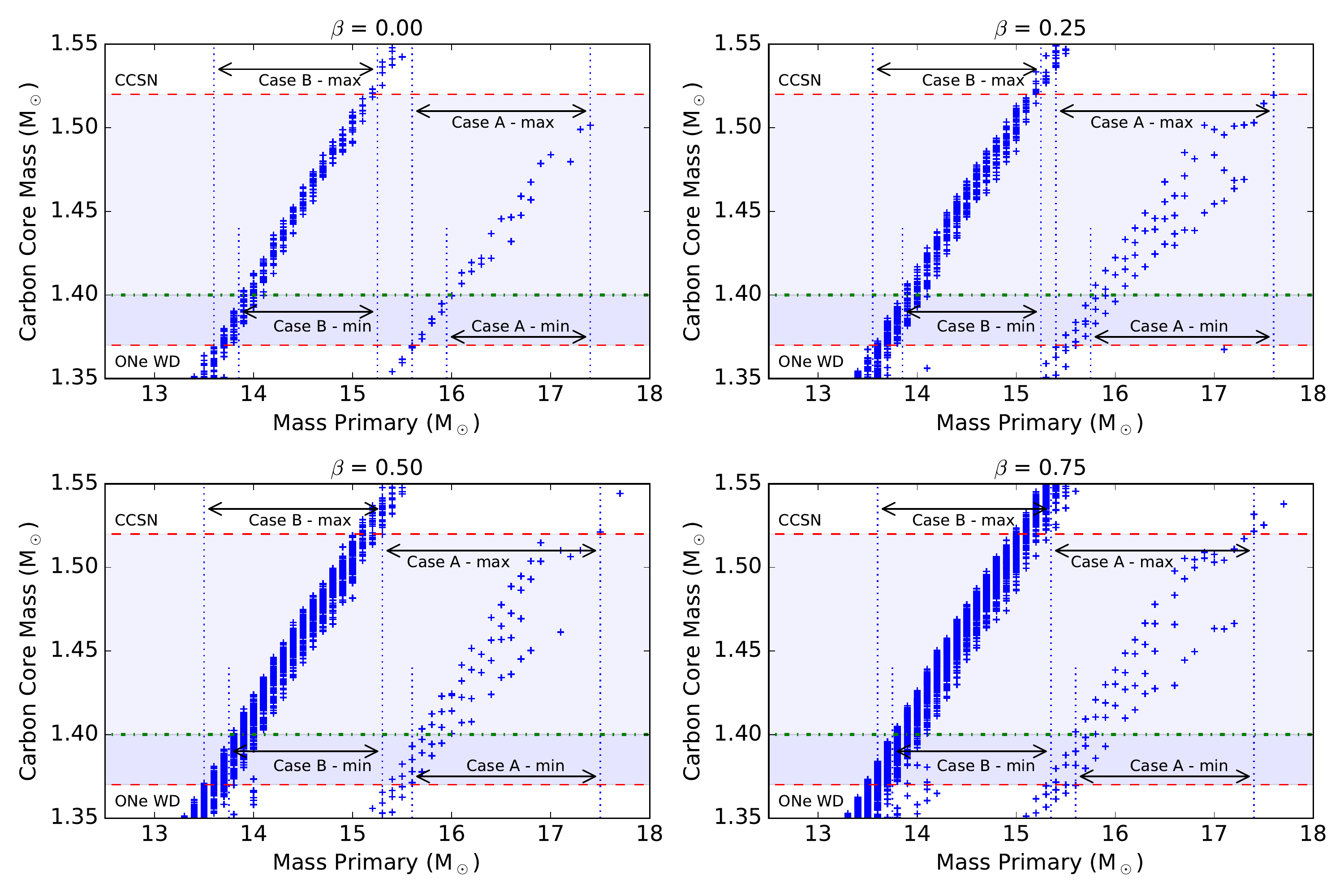}
\caption{Final $M_\mathrm{CO}$ as a function of initial mass, organized by mass transfer efficiency ($\beta$) in four panels. The horizontal dashed red at $1.52 \msun$ indicates the \replaced{highest}{maximum} possible $M_\mathrm{CO}$ that does not lead to off-center neon ignition, and thus forms the upper boundary of the ECSN range. The horizontal dash-dotted green line at $1.4\msun$ indicates the effective Chandrasekhar mass. The horizontal dashed red line at $1.37 \msun$ indicates the lowest possible CO core mass that might be able to evolve into an ECSN. The light shaded area between $1.37 \msun \le M_\mathrm{CO} \le 1.52 \msun$ and $1.4\msun \le M_\mathrm{CO} \le 1.52 \msun$ indicates the maximum and most feasible ECSN range, as long as the decrease in temperature is stopped and the density is able to increase again. \deleted{Combine this region with the darker blue shaded area between $1.37\msun$ and $1.4\msun$ and the maximum CO mass range for ECSN is obtained. This translates to minimum and maximum initial masses for ECSN based on the various options (horizontal lines), and subsequently various possible mass ranges, indicated by arrows, for both Case A mass transfer and Case B mass transfer. The minimum and maximum mass ranges for Case A and Case B mass transfer are only marginally affected by the mass transfer efficiency $\beta$. Since our grid has a $0.5$ day spacing in the period range of Case A mass transfer, we possibly under-sampled the Case A channel as only the models with an initial period of $3$ days led to Case A systems. A finer grid, which samples the full period range of Case A systems could lead to a Case A channel that is wider in terms of $M_\mathrm{CO}$.}}
\label{fig:ECSN_range}
\end{center}
\end{figure*}

\replaced{If we assume the gloom picture sketched above, the mass range for ECSN from close binary systems is non-existent (at least in the direct scenario; see for an indirect scenario Section~\ref{sec:indirect_scenario}). If we, however, }{If we} assume that continued core contraction after mass loss ceases will eventually converge the evolution a core onto a common evolutionary track in the $\rho_\mathrm{c}-T_\mathrm{c}$ plane toward electron captures, \replaced{a much wider initial mass range opens up as possible ECSN candidates}{we are able to characterize, within the limits of our assumptions on the treatment of convection, overshooting, mixing, and accretion, the initial mass range for ECSN}. Figure~\ref{fig:ECSN_range} provides a different way of looking at our dataset, showing the final $M_\mathrm{CO}$ as a function of the initial primary mass. The horizontal dashed red \added{line} at $1.52 \msun$ indicates the \replaced{lowest possible}{maximum} $M_\mathrm{CO}$ that \replaced{allows for off-center}{avoids} neon ignition, and hence forms the upper boundary of the ECSN range. This upper boundary could be shifted to even higher $M_\mathrm{CO}$ if neon flames are able to stall \citep{jhn14}, leaving the  \replaced{central conditions}{chemical composition in the center} unaltered and allowing for subsequent electron captures to occur. Most work on stalling flames has been done in the context of carbon flames by \citet{dht+13, dth+15, fft15}  whose research suggest that efficient convective boundary mixing can disrupt inwardly propagating carbon flames, however \citet{lsq+16} find that hybrid cores are unlikely. \replaced{As quenched neon flames are not the topic of this paper (see for more details on neon flames \citealt{jhn14}), we recognize the possibility that our upper boundary might have to be modified when the details become more clear.}{We defer to future work the investigation of quenched neon flames which might influence our upper boundary for $M_\mathrm{CO}$ leading to ECSN.} The horizontal dash-dotted green line at $1.4\msun$ indicates the Chandrasekhar mass for a composition of 50\% neon and 50\% oxygen. The horizontal dashed red line at $1.37 \msun$ indicates the lowest possible CO core mass that is able to evolve into an ECSN. 

Two distinct sequences are visible in each panel \citep[cf.][Fig 5b]{wl99b}, Case B models (on the left) and Case A models (on the right). The reason that Case B and Case A systems are separate is that mass loss during the main sequence reduces the hydrogen burning core, effectively shifting the ECSN range \added{for Case A systems} to higher initial masses. The scatter in the vertical direction can be attributed to a spread in initial period, with models toward the lower end of the sequence originating from low period binaries, while models toward the higher end of the sequence originating from higher period binaries, giving rise to smaller, resp. larger CO cores (c.f. the discussion in Section~\ref{sec:rep_systems}, see also \citealt{wlb01}). As argued \replaced{before (}{in} Sec.~\ref{sec:ECSN_mass_range}, initial periods that are \replaced{greater}{longer than our ECSN candidates}\deleted{and smaller than our ECSN candidates shown in Fig.~\ref{fig:ECSN_panel}} lead to contact systems\added{, while initial periods that are shorter than our ECSN candidates lead to either a CO core below the critical mass for neon ignition, or to a Case A contact system}. Thus it appears that for a given initial primary mass the initial period will allow for a certain spread in final $M_\mathrm{CO}$. However, the key factor that determines the location of the ECSN channel is the initial primary mass. While the number of models that are able to produce CO cores that are favorable for the development ECSN increases with increasing $\beta$ the ECSN mass range itself does not really shift. This is true for both the Case A ECSN channel as \replaced{well as}{for} the Case B ECSN channel. If we allow all \added{CO} cores with $1.37 \msun \le M_\mathrm{CO} \le1.52 \msun$ to eventually explode as an ECSN we find an initial mass range for ECSN between $13.5 \msun$ and $15.25 \msun$ for Case B systems and between $15.4 \msun$ and $17.6 \msun$ for Case A systems. This mass range is indicated in Figure~\ref{fig:ECSN_range} with arrows marked ``Case B -- max'' and ``Case A -- max'', respectively. If we confine the cores that are able to evolve into an ECSN to $M_\mathrm{CO}$ between the Chandrasekhar mass ($1.40 \msun$) and $1.52 \msun$ the initial mass range for ECSN narrows to $14.2 \msun$ and $15.25 \msun$ for Case B systems and between $16.1 \msun$ and $17.6 \msun$ for Case A systems. This mass range is indicated in Figure~\ref{fig:ECSN_range} with arrows marked ``Case B -- min'' and ``Case A -- min'', respectively. \added{Both channels are slightly offset from the single star ECSN channel, which, for the chosen input physics and convection criterion, is located between $12.7 \msun$ and $13.2 \msun$, based on  $1.38 \msun \le M_\mathrm{CO, single} \le 1.42 \msun$. This difference in initial masses is small and probably difficult to detect observationally in a population consisting of both single and binary stars.} As our input physics in terms of convective mixing is similar to \citet{wlb01}, i.e. we don't take into account additional mixing such as convective overshooting or exponentially decreasing diffusion \cite{her00}, we expect that this initial mass range might shift to lower initial masses by $2-3 \msun$ if additional mixing is taken into account. It should, however, not have significant effects on the evolution or the structure of the models or any of our further conclusions (see App.~\ref{sec:other_mixing}).

It is worth noting that more ECSN are predicted for systems with a mass ratio close to unity, as the development of contact happens at \replaced{larger}{longer} period for higher $q$ systems (see Fig.~\ref{fig:ECSN_panel}). As the primary star starts transferring mass to the secondary, the orbit shrinks, until the mass ratio is reversed. This reversal happens earlier for mass ratios close to one, and later for lower mass ratios, increasing the chance for contact \citep{mli+13}. This primarily affects the numbers of ECSN, not so much the initial mass range (except for Case A systems). This is also true for the value of $\beta$, which controls the amount of matter that is lost from the system (i.e., $\beta = 0$ is conservative mass transfer, no mass that is transferred from the primary to the secondary is lost from the system; $\beta = 1$ is completely non-conservative mass transfer, all mass that is transferred from the primary is lost from the system, no accretion onto the secondary). For the sake of our parameter study we chose various fixed values of $\beta$, while the mass transfer efficiency in real systems \added{varies in time and} will depend on the evolutionary phase of both stars, the amount of matter already accreted onto the secondary and how that has affected its spinrate. Several mechanisms have been suggested \replaced{that}{which} \replaced{regulate}{control} the efficiency of mass transfer, mass accretion and mass loss, including the existence of an accretion disk that regulates the amount of mass and angular momentum that can be accreted \citep{pac91, pn91, dsd+13}, the necessity of the secondary to stay below critical rotation \citep{pac81}, and the effects of tides on the stellar spins and the stellar orbit \citep{zah77,htp02}. Work by \citet{dsd+13, rgl+08} for systems with slightly lower masses suggests periods with values of $\beta$ close to $1$ (i.e. very inefficient mass transfer), while simulations with a strong tidal interaction (i.e. a short spin-orbit synchronization timescale) suggest shorter periods of moderately inefficient mass transfer \citep{mesa3}. Although our models suggest that the efficiency of mass transfer does not really affect the mass range for ECSN, it does, however, strongly affect the range of initial periods that can \replaced{give rise to}{lead to} an ECSN. In addition, the evolution of the secondary will be affected. It will most likely rapidly spin up after the onset of mass transfer and maintain near-critical rotation for possibly extended periods of time. This will induce strong rotational mixing \citep{lan12, mcl+08, mli+13}, causing possibly quasi-chemically homogeneous evolution \citep{med87, lan12} and alter the evolution of the star beyond just the simple fact of mass accretion \citep{hmm04c, mm16, mlp+16}. Although it is not clear to what extent this will affect the incidence of ECSN, the effects of tides, mass and angular momentum transfer and loss, and near-critical rotation of the secondary are possibly important and will be discussed in a forthcoming paper, in addition to the effects of additional mixing and convection criteria. 

If the scenario \added{for the formation of ECSN in binaries, as discussed in this paper,} turns out to \replaced{succeed}{correct}, the binary ECSN channel \replaced{will}{might} contribute to the population of NS+NS systems that can be observed with aLIGO/Virgo \citep{cbf+17}. Precursors to these NS+NS systems will be visible as Be/X-ray binaries \citep{kcp11,sl14}. \added{With the discovery of GW170817 \citep{ligo-gw170817}, electron-capture supernovae have been propelled into the spotlight \citep{ligo17}. As the progenitors of NS+NS collisions have to survive two supernova explosions, while still remain bound, ECSN are promising candidates as the NS kick is expected be much smaller for ECSN than for CCSN \citep{plp+04}. While our research has focused on the first supernova in a binary system, low natal kicks from ECSN are particularly relevant in the second explosion \citep{tkf+17}. However, our conclusion that CO cores with masses up to $1.52 \msun$ can produce ECSN is potentially relevant in determining the final fate of so-called ultra-stripped supernovae, the second supernova in a NS+NS system \citep{tlm+13, tlp15, tkf+17}.}

\acknowledgments
We thank the anonymous referee for very detailed and constructive feedback, which significantly contributed to improving the quality of this paper. We thank Neile Havens for excellent technical support and Harvard Townsend for administrative IT support. AJP acknowledges support from the Aldeen Fund. SW, SH, and JT like to thank Dr. Dorothy Chappell, Dean of Natural and Social Sciences at Wheaton College and the Wheaton College Alumni Association for support for the summer research program. We thank the MESA council for making available, maintaining and continuously developing an outstanding stellar evolution code. This research used the High Performance Computing resources at Wheaton College and also has made use of NASA's Astrophysics Data System.

\software{MESA (v8118; \citealt{mesa1, mesa2, mesa3}), pyMESA, \citealt{far17}}.

\bibliographystyle{apj}


\appendix
\section{Overshooting, Fast Semi-Convection and the Critical $M_\mathrm{CO}$ for Neon Ignition}\label{sec:other_mixing}
The models presented in this paper use the Ledoux criterion for determining convective boundaries, and a small amount of semi-convection, $\alpha_\mathrm{sm} = 0.01$. In order to investigate the robustness of our results against differences in the treatment of convection, we calculated two sequences (similar to the ones presented in Sec.~\ref{sec:late_mass_loss}), one with a stronger semi-convection ($\alpha_\mathrm{sm} = 1.0$) and the other using overshooting ($f_\mathrm{ov} = 0.016$, in the context of the Schwarzschild criterion). In this way we are able to determine whether the critical mass for neon ignition (in this paper found to be $M_\mathrm{CO} \approx 1.52 \msun$) shifts to higher or lower $M_\mathrm{CO}$ with different convection criteria.

\begin{figure*}[htbp]
\begin{center}
\includegraphics[scale=0.58]{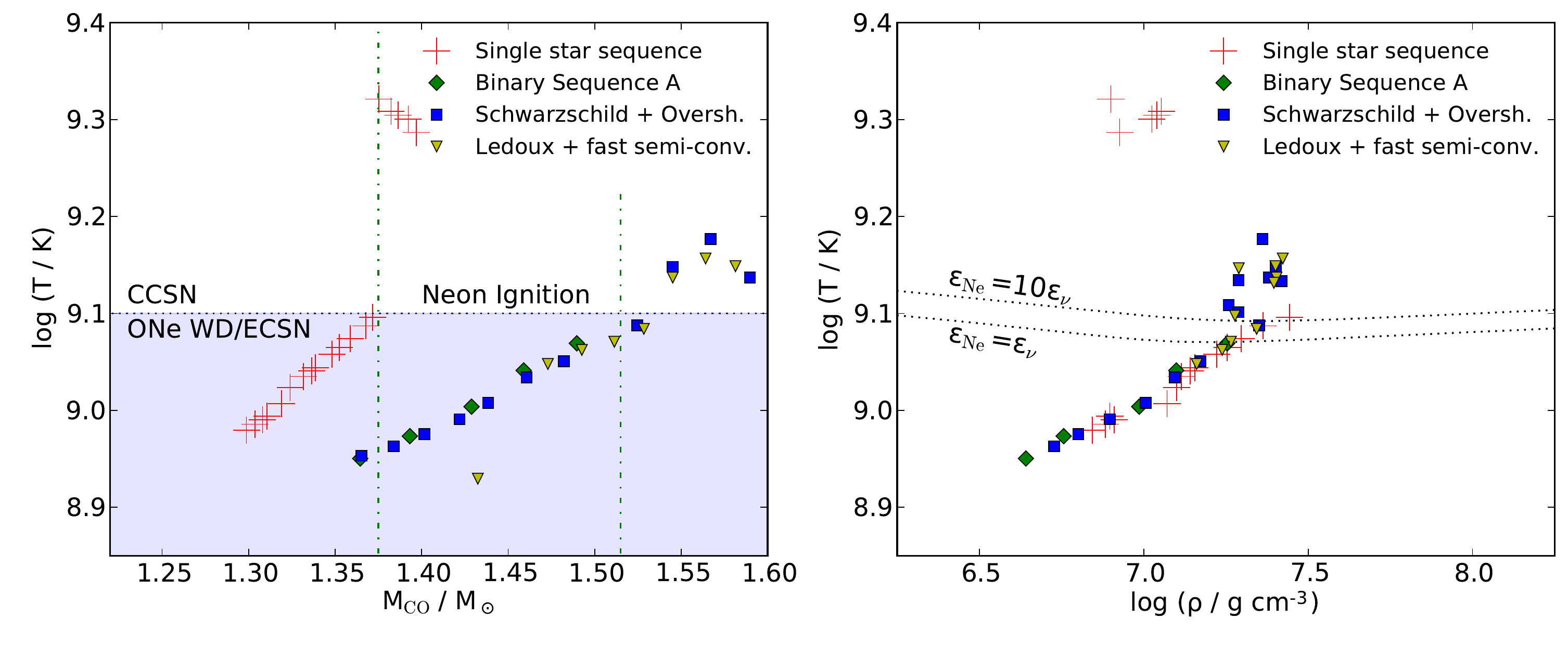}
\caption{Maximum temperature attained in the core as a function of $M_\mathrm{CO}$ (left panel) and as a function of the density at the mass coordinate of $T_\mathrm{max}$ at the time of $T_\mathrm{max}$ (right panel), for single stars (red crosses) and primary stars in a binary system (filled symbols) with various treatments of convection and overshooting. The dotted line in the left panel approximates the critical temperature for neon ignition, while the dotted lines in the right panel give the locations where the energy generation of neon burning is equal to, and 10 times exceeds the energy losses due to thermal neutrinos.}
\label{fig:mcore_maxT_other}
\end{center}
\end{figure*}

Figure~\ref{fig:mcore_maxT_other} shows these various sequences (including Binary Sequence A from Fig.~\ref{fig:mcore_maxT}), calculated with different treatments of convection. Each sequence was calculated with initial masses between $10.0 \msun$ and $15.5 \msun$, for a mass ratio of $q=0.8$, a period of $10$ days and a mass transfer efficiency, $\beta = 0.5$. The sequence with $\alpha_\mathrm{sm} = 1.0$ is shown with yellow symbols, and the sequence with Schwarzschild with $f_\mathrm{ov} = 0.016$ with blue symbols.

While we defer a full investigation to a future work, this initial investigation shows that each of these modifications results in a relationship between $M_\mathrm{CO}$ and $\log\ T_\mathrm{max}$ that is similar to the one shown in Fig.~\ref{fig:mcore_maxT}. This confirms that the critical mass for neon ignition in binary stars is not dependent on the adopted convection criterion, the efficiency of semi-convection, or the use of overshooting. However, each of these different modification does affect the initial mass range by shifting it to lower initial masses (c.f. \citealt{phl+07, dsl+10}).

\section{The Effects of Variation of the Spatial and Temporal Resolution}\label{sec:resolution}
\begin{figure*}[tbp]
\begin{center}
\includegraphics[scale=0.58]{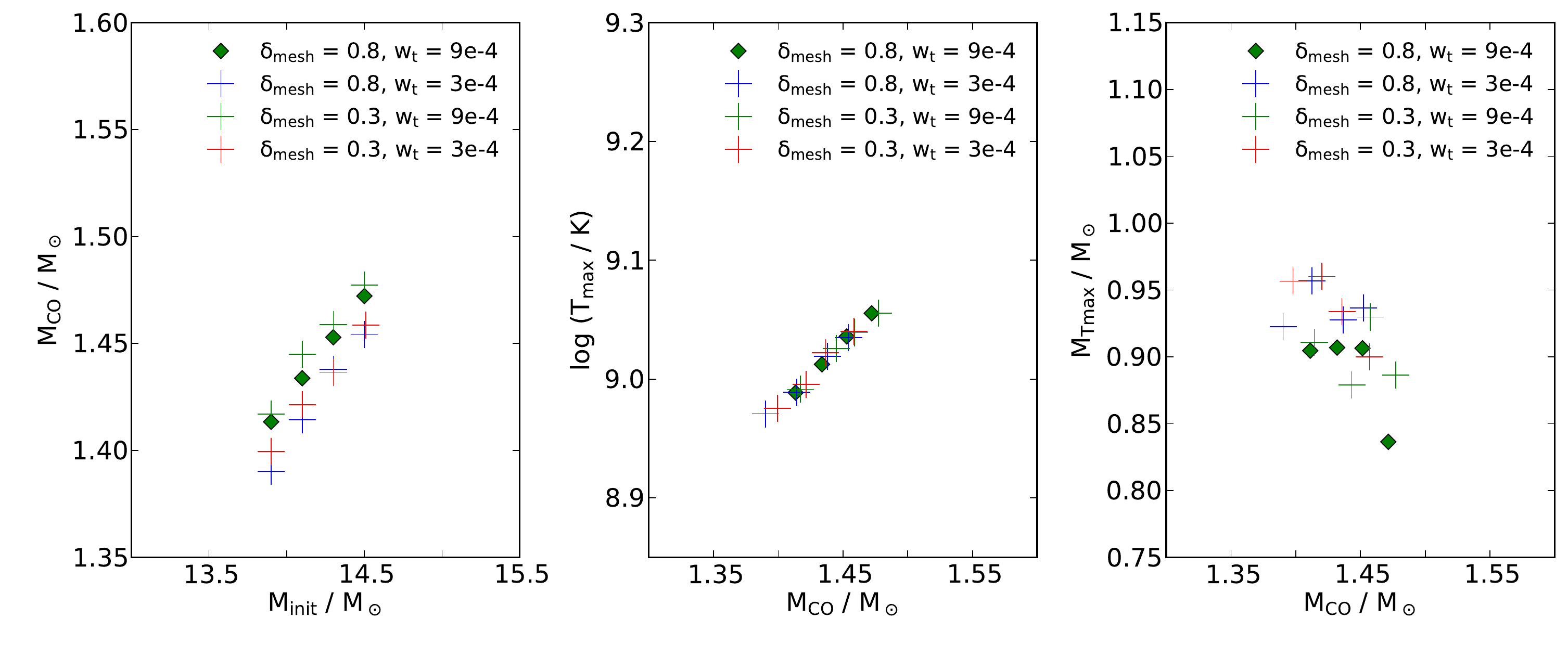}
\caption{Results of our resolution study. Our baseline parameters are shown as green diamonds while variations in $w_\mathrm{t}$ and/or $\delta_\mathrm{mesh}$ are shown with colored plus signs. The left panel shows the initial primary mass versus final CO core mass. The middle panel shows the CO core mass versus $T_\mathrm{max}$, and the right panel shows the CO core mass versus the radial mass coordinate of $T_\mathrm{max}$. All panel shows that our baseline models are within the convergence envelope and thus producing reliable results.}
\label{fig:resolution}
\end{center}
\end{figure*}

As stellar models can potentially be quite sensitive to the spatial and temporal resolution, we conduct a small-grid resolution study to investigate the robustness of our models against spatial and temporal variations in resolution. While MESA has many parameters that control the temporal and spatial resolution in detail, the parameter $w_\mathrm{t}$ controls the overall temporal resolution, and the parameter $\delta_\mathrm{mesh}$ controls the overall spatial resolution. The calculations described in this paper were computed with our baseline parameters $w_\mathrm{t} = 9 \times 10^{-4}$ and $\delta_\mathrm{mesh} = 0.8$. To investigate convergence of our models at a different resolution, we computed a sequence with $w_\mathrm{t} = 9 \times 10^{-4}$ and $\delta_\mathrm{mesh} = 0.3$ (increased spatial resolution), $w_\mathrm{t} = 3 \times 10^{-4}$ and $\delta_\mathrm{mesh} = 0.8$ (increased temporal resolution) and $w_\mathrm{t} = 3 \times 10^{-4}$ and $\delta_\mathrm{mesh} = 0.3$ (increased spatial and temporal resolution). The results are plotted in Fig.~\ref{fig:resolution}. The left panel shows the initial mass of the star versus the final $M_\mathrm{CO}$, comparable to Fig.~\ref{fig:ECSN_range}. The models with the baseline values are shown as green diamonds, while the variations in $w_\mathrm{t}$ and $\delta_\mathrm{mesh}$ are shown with plus signs. A minimal spread can be seen in the resulting $M_\mathrm{CO}$, with models with $w_\mathrm{t} = 9 \times 10^{-4}$ slightly more massive than models with $w_\mathrm{t} = 3 \times 10^{-4}$. The middle panel shows the  maximum temperature attained in the core, plotted against the CO core mass (similar to Fig.~\ref{fig:mcore_maxT}). This panel shows excellent model convergence in the mass range considered, with no impact due to changes in spatial or temporal resolution. The location of $T_\mathrm{max}$ (shown in the far right panel) is a bit more sensitive to changes in the spatial and temporal resolution, but this can also be attributed to a steep relationship between $M_\mathrm{CO}$ and $M_\mathrm{Tmax}$ (c.f. Fig.~C2 in \citealt{sqk16}). However, this variation has no impact on any of our results.

Based on this resolution study we conclude that our baseline parameters lead to good model convergence, and are comparable to models with a higher spatial and/or temporal resolution. The most notable difference is that our baseline models produce more massive cores than models with a higher spatial and temporal resolution. While this does not affect the relationship between the maximum temperature attained in the core and the CO core mass, a higher resolution grid would shift the initial mass range to higher initial primary masses.

\listofchanges
\end{document}